\documentclass[10pt,conference]{IEEEtran}
\IEEEoverridecommandlockouts

\usepackage{tikz}
\usetikzlibrary{quantikz2}

\usepackage[T1]{fontenc}

\usepackage{cite}
\usepackage{amsmath}
\usepackage{amssymb}
\usepackage{amsfonts}
\usepackage{amsthm}
\usepackage{dsfont}
\usepackage{algorithmic}
\usepackage{graphicx}
\usepackage{textcomp}
\def\BibTeX{{\rm B\kern-.05em{\sc i\kern-.025em b}\kern-.08em
    T\kern-.1667em\lower.7ex\hbox{E}\kern-.125emX}}

\usepackage{placeins}
\usepackage{booktabs}
\usepackage{multirow}

\usepackage{subcaption}

\PassOptionsToPackage{hyphens}{url}\usepackage{hyperref}
\usepackage{xspace}

\newcommand{\C}{\mathbb{C}}
\newcommand{\R}{\mathbb{R}}

\renewcommand{\Im}[1]{\text{Im}\left( #1 \right)}
\renewcommand{\Re}[1]{\text{Re}\left( #1 \right)}
\newcommand{\pars}[1]{\left( #1 \right)}

\newcommand\norm[1]{\left\lVert#1\right\rVert}
\newcommand\abs[1]{\left\lvert#1\right\rvert}

\theoremstyle{plain}
\newtheorem{proposition}{Proposition}[subsection]

\everymath{\small}

\begin{document}

\title{An Empirical Analysis on the Effectiveness of the Variational Quantum Linear Solver}

\author{
\IEEEauthorblockN{Gloria Turati}
\IEEEauthorblockA{\textit{Politecnico di Milano}\\
Milano, Italy\\
\href{mailto:gloria.turati@polimi.it}{gloria.turati@polimi.it}}
\and
\IEEEauthorblockN{Alessia Marruzzo}
\IEEEauthorblockA{\textit{Eni SpA}\\
San Donato Milanese, Italy\\
\href{mailto:alessia.marruzzo@eni.com}{alessia.marruzzo@eni.com}}
\and
\IEEEauthorblockN{Maurizio Ferrari Dacrema}
\IEEEauthorblockA{\textit{Politecnico di Milano}\\
Milano, Italy\\
\href{mailto:maurizio.ferrari@polimi.it}{maurizio.ferrari@polimi.it}}
\and
\IEEEauthorblockN{Paolo Cremonesi}
\IEEEauthorblockA{\textit{Politecnico di Milano}\\
Milano, Italy\\
\href{mailto:paolo.cremonesi@polimi.it}{paolo.cremonesi@polimi.it}}
}

\maketitle

\thispagestyle{plain}
\pagestyle{plain}

\begin{abstract}
Variational Quantum Algorithms (VQAs) have emerged as promising methods for tackling complex problems on near-term quantum devices. Among these algorithms, the Variational Quantum Linear Solver (VQLS) addresses linear systems of the form $Ax=b$, aiming to prepare a quantum state \ket{x} such that $A$\ket{x} is proportional to the quantum state corresponding to $b$. A key advantage of VQLS is its use of amplitude encoding, which requires a number of qubits that scales logarithmically with the linear system size.
However, the existing literature has primarily focused on linear systems of limited size or with a specific structure. In this study, we extend the application of VQLS to more general and larger problem instances, including problems where state preparation is non-trivial and problems within the real domain of fluid dynamics.
Our investigation reveals some critical challenges inherent to VQLS, including the need for a sufficiently expressive ansatz, the large number of circuit executions required to estimate the cost function, and the high gate count in the circuits in the most general setting.
Our analysis highlights the obstacles that need to be addressed for a broader application of VQLS and concludes that further research is necessary to fully leverage the algorithm's capabilities in addressing real-world problems.
\end{abstract}

\begin{IEEEkeywords}
Quantum Algorithms, VQAs, VQLS
\end{IEEEkeywords}

\section{Introduction}

Quantum computing has emerged as a revolutionary paradigm in computer science, showing the potential to accelerate classical computation\cite{preskill_2018, arute_2019, kim_2023, sood_2023, daley_2022}. However, the current limitations of quantum technology, such as the restricted number of qubits and inherent noise in the devices, pose significant challenges to their widespread adoption\cite{franca_2021, pan_2023, fellous-asiani_2021, clerk_2010}.

Variational Quantum Algorithms (VQAs)\cite{cerezo_2021_vqas} have emerged as a class of algorithms designed to overcome these limitations for a variety of problems. VQAs leverage a parametric quantum circuit (ansatz) in conjunction with a classical optimizer that adjusts its parameters.
The parameter optimization process minimizes a cost function such that running the quantum circuit with the optimal parameters should yield a good solution to the problem.
VQAs offer several advantages, including shallower circuits and inherent error mitigation, which make them well-suited for the currently available Noisy Intermediate-Scale Quantum (NISQ)\cite{preskill_2018} devices.
However, the efficacy of VQAs is strictly tied to the choice of the ansatz and the cost function \cite{qin_2023, wurtz_2021, sim_2019, du_2020}. In particular, a well-known issue for VQAs are barren plateaus\cite{mcclean_2018, arrasmith_2021, arrasmith_2022, cerezo_2021, holmes_2022, larocca_2022, volkoff_2021, sanavio_2024}, where the gradient exponentially vanishes with the circuit size, which makes challenging for the classical optimizer to find the global optimum. To mitigate this issue, it becomes crucial to define an appropriate cost function and to select an efficient optimizer.

Well known VQAs include the Variational Quantum Eigensolver (VQE)\cite{peruzzo_2014, tilly_2021}, widely applied in quantum chemistry \cite{cao_2019, mcclean_2016}, and the Quantum Approximate Optimization Algorithm (QAOA)\cite{farhi_2014, blekos_2023}, extensively employed for solving combinatorial optimization problems, including MaxCut \cite{crooks_2018, wang_2018, basso_2022, farhi_2017, shaydulin_2023, stechly_2023, zhou_2020, larkin_2022}, Minimum Vertex Cover \cite{cook_2020}, Graph Coloring \cite{tabi_2020_graph_colouring}, Constraint Satisfaction \cite{lin_2016_csp, willsch_2020}, and more complex problems like Prime Factorization \cite{yan_2022_factoring}, the Traveling Salesman problem \cite{radzihovsky_2019_tsp}, Portfolio Optimization \cite{brandhofer_2022_portfolio_opt}, and the Job Shop Scheduling problem \cite{kurowski_2023_jssp}.

The Variational Quantum Linear Solver (VQLS), introduced by Bravo-Prieto et al. \cite{bravo-prieto_2023}, is a VQA specifically designed to solve systems of linear equations with quantum computers. VQLS offers a NISQ-friendly alternative to the Harrow-Hassidim-Lloyd (HHL) algorithm\cite{harrow_2009_hhl, zaman_2023_hhl}, which also aims to solve linear systems and claims exponential speedup over classical algorithms. HHL, indeed, relies on deep circuits that are more susceptible to noise, which makes its implementation on the currently available quantum devices impractical.
In contrast, VQLS leverages shallower circuits, making the algorithm compatible with NISQ devices.
VQLS relies on amplitude encoding\cite{schuld_2018}, which allows the solution of the linear system to be represented using a number of qubits which is logarithmic in the system size, potentially leading to a quantum advantage.

Several studies have explored the potential of VQLS in diverse scenarios: some works have focused on finding applications of VQLS across various problem domains \cite{liu_2024, trahan_2023, liu_2022, shang_2023, luo_2024, xing_2023, ali_2023}, whereas others have proposed modifications to the algorithm itself \cite{patil_2022, huang_2021, yi_2023, pellow-jarman_2023, saito_2023} or benchmarked its performance using different optimizers \cite{pellow-jarman_2021}.
However, these works predominantly focus on small-scale problem instances or instances characterized by specific structures, where performing the LCU decomposition is straightforward and results in gates easily implementable on the quantum circuit.
Therefore, our research aims to build upon these studies and evaluate the effectiveness of the algorithm on a broader class of problems. Specifically, our contribution includes:
\begin{itemize}
    \item investigating the VQLS effectiveness on problem instances characterized by larger and more generally structured system matrices;
    \item examining the required circuit depth for implementing VQLS, with a specific focus on scenarios where the state preparation component is non-trivial;
    \item proposing an expression for the cost function, which utilizes a quarter of the number of terms found in the existing literature (to the best of our knowledge), thereby reducing the required number of circuit executions.
\end{itemize}
Our overarching goal is to unveil strengths and limitations of VQLS, providing a comprehensive understanding of its effectiveness.

\section{The Variational Quantum Linear Solver}

\subsection{Algorithm overview}

The Variational Quantum Linear Solver \cite{bravo-prieto_2023} (VQLS) considers a system of linear equations in the form $Ax = b$, where $A\in\C^{N \times N}$ is the system matrix and $b\in\C^N$ is a vector. The system size $N$ is specifically chosen to be a power of 2. The outcome of VQLS is a quantum state \ket{x} such that $A \ket{x}$ is proportional to \ket{b}, a normalized version of $b$.\footnote{This is equivalent to preparing a quantum state $\ket{x}$ that is proportional to a vector $x$ satisfying the equation $Ax = b$.}
The algorithm requires the system matrix $A$ to be expressed through a linear combination of unitary matrices, known as LCU decomposition (see Section \ref{LCU_decomposition_subsec}). The selection of these unitary matrices plays a pivotal role in exploiting the full capabilities of VQLS and achieving optimal performance.
Furthermore, the algorithm requires to prepare state \ket{b}, which is an important assumption since finding the appropriate sequence of gates able to prepare an arbitrary quantum state is often not straightforward.

The solution to the linear system is encoded in the amplitudes of the final state \ket{x} of the quantum circuit utilizing amplitude encoding \cite{schuld_2018}. 
The circuit for producing this solution is determined through a variational approach, where a quantum circuit is established and its parameters are optimized using a classical optimizer.

The operator representing the ansatz is denoted by $V(\theta)$, where $\theta$ is a real-valued parameter vector, whose size depends on the number of parameters in the specific ansatz employed. 
The quantum state produced by the ansatz, denoted as \ket{x(\theta)}, is generated by applying the operator $V(\theta)$ on the initial state \ket{0}, as in
\begin{equation}
    \ket{x(\theta)} = V(\theta) \ket{0}.
\end{equation}
The aim is to identify the optimal parameter vector, $\theta_{\text{opt}}$, such that the state $\ket{x(\theta_{\text{opt}})}$ closely approximates the solution to the linear system. Ideally, $A\ket{x(\theta_{\text{opt}})}$ should be proportional to \ket{b}.

This is achieved by defining a cost function $C(\theta)$, which quantifies the deviation of $A\ket{x(\theta)}$ from being proportional to \ket{b} (see Section \ref{cost_function_subsec}).
The value of $C(\theta)$, for a given set of parameters, is estimated by executing numerous quantum circuits, which depend on the matrix $A$ and the vector $b$.
After the cost function value for a particular $\theta$ is estimated, a classical optimizer adjusts the parameters and a new iteration of the algorithm begins. This process continues until a predefined termination condition is satisfied.
Once the optimal parameter vector $\theta_{\text{opt}}$ has been found, the solution of the problem is the state $\ket{x(\theta_{\text{opt}})} = V(\theta_{\text{opt}}) \ket{0}$.

\subsection{LCU Decomposition} \label{LCU_decomposition_subsec}
An essential prerequisite for the application of VQLS is to decompose the matrix $A$ into a \textit{Linear Combination of Unitaries}, commonly referred to as LCU decomposition:
\begin{equation}
    \label{eq:LCU_decomposition}
    A = \sum_{i=0}^{L-1} c_i A_i,
\end{equation}
where $A_i$ are unitary matrices. 
A widely employed technique for performing such a decomposition for a matrix $A$ of size $N=2^n$ involves considering a basis of matrices $A_i$ constructed using the tensor product of Pauli matrices:
\begin{equation}
    \label{eq:pauli_LCU}
    \mathcal{B} = \left\{A_i = \prod_{j=0}^{n-1} S_j \quad \forall S_j \in \{I, X, Y, Z\} \right\}.
\end{equation}
For each matrix $A_i$, the corresponding coefficient is given by
\begin{equation}
    \label{eq:pauli_ci}
    c_i = \frac{\text{Tr}(A A_i)}{2^n}.
\end{equation}

An important advantage of utilizing Pauli matrices for the decomposition is that they are easy to implement on the quantum hardware.
However, a notable limitation arises from the fact that in the most general case this basis contains $L = 4^n = N^2$ matrices and, as such, cancels any performance improvement derived from the amplitude encoding of \ket{x}. While there are instances, such as the \textit{ising} problem discussed in Section \ref{sec:experimental_protocol}, where the required number of matrices is lower, this is not the general scenario. 
Since the number of circuits required for estimating the value of the cost function is $O(L^2)$ (see Appendix \ref{circuit_execution_appendix}), a decomposition with $L=N^2$ implies that the number of circuit evaluations required by the algorithm is $O(N^4)$. Considering that solving a system of linear equations with the classical Gauss Elimination method requires $O(N^3)$, it is clear that using a more efficient LCU decomposition is a crucial and open problem for VQLS.
Although alternative techniques for the LCU decomposition \cite{bravo-prieto_2023, liu_2021, xu_2021, trahan_2023, cappanera_2021} do exist, with some, like the approach presented in \cite{bravo-prieto_2023}, employing as few as 4 unitaries, these methods are tailored to specific classes of problems or require a significantly higher number of gates for implementing the unitary matrices. Consequently, adopting these techniques becomes impractical on real quantum hardware. 

\subsection{Cost function} \label{cost_function_subsec}

There are two versions of the VQLS cost function described in the original paper \cite{bravo-prieto_2023}: \textit{global} and \textit{local}.
To simplify the notation, we will use \ket{\psi} to denote $A\ket{x(\theta)}$.

\paragraph{Global Cost Function}

The global cost function is defined as
\begin{equation}
    \label{eq:global_cost_function}
    C_G(\theta) 
    = \frac{\hat{C_G}(\theta)}{\braket{\psi}{\psi}}
    = 1 - \frac{\abs{\braket{b}{\psi}}^2}{\braket{\psi}{\psi}},
\end{equation}
where $\hat{C_G}(\theta) = \bra{x(\theta)}H_G\ket{x(\theta)}$, $H_G = A^\dag (\mathds{1} - \ket{b} \bra{b}) A$, and $\mathds{1}$ is the $N$-dimensional identity matrix.

Therefore, for each given $\theta$, the computation of the cost function involves estimating the quantities $\braket{\psi}{\psi}$ and $\abs{\braket{b}{\psi}}^2$.
In particular, if we choose a Pauli basis (\ref{eq:pauli_LCU}) to perform the LCU decomposition (\ref{eq:LCU_decomposition}) as described in Section \ref{LCU_decomposition_subsec}, and the matrix $A$ is Hermitian, then we can compute $\braket{\psi}{\psi}$ and $\abs{\braket{b}{\psi}}^2$ using the following expressions:
\begin{equation}
    \label{eq:real_glob_den}
    \begin{split}
        \braket{\psi}{\psi}
        &= \sum_{i=0}^{L-1} c_i^2 
        + \sum_{i=0}^{L-1} \sum_{j=i+1}^{L-1} 2 c_i c_j \Re{\bra{0} V^\dag A_i^\dag A_j V \ket{0}}
    \end{split}
\end{equation}
\begin{equation}
    \label{eq:real_glob_num}
    \begin{split}
        &\abs{\braket{b}{\psi}}^2
        = \sum_{i=0}^{L-1} c_i^2 \abs{\bra{0} U^\dag A_i V \ket{0}}^2 \\
        &+ \sum_{i=0}^{L-1} \sum_{j=i+1}^{L-1} 2 c_i c_j \Re{\bra{0} U^\dag A_i V \ket{0}} \Re{\bra{0} U^\dag A_j V \ket{0}} \\
        &+ \sum_{i=0}^{L-1} \sum_{j=i+1}^{L-1} 2 c_i c_j \Im{\bra{0} U^\dag A_i V \ket{0}} \Im{\bra{0} U^\dag A_j V \ket{0}}. \\
    \end{split}
\end{equation}
The equations are summations of terms that can be computed employing Hadamard tests (see Appendix \ref{hadamard_test_appendix}). 

Previously, the required number of Hadamard tests for computing the denominator of the global cost function in (\ref{eq:global_cost_function}) was $2L^2$ ($L^2$ for the real parts of all the expectations, $L^2$ for the imaginary parts), and additional $2L$ Hadamard tests were necessary for computing the numerator, for a total of $2L^2 + 2L$ Hadamard tests for estimating the global cost function. By using (\ref{eq:real_glob_den}), we only require $\frac{L(L-1)}{2}$ Hadamard tests for estimating the denominator and, using (\ref{eq:real_glob_num}), we still need $2L$ Hadamard tests for the numerator, resulting in a total of $\frac{L^2 + 3L}{2}$ Hadamard test executions, which is asymptotically a quarter of the previously required number.
Furthermore, if we adopt a Pauli LCU decomposition, we find that in the most general case, $L = N^2$, with $N$ being the system size. Therefore, in terms of the system size, previously we needed to perform $2N^4 + 2N^2$ Hadamard tests, while now we require only $\frac{N^4 + 3N^2}{2}$ executions.
We report the details of our derivation as well as instructions on how to apply the Hadamard test in Appendix \ref{cost_function_appendix} and \ref{circuit_execution_appendix}.

\paragraph{Local Cost Function}

The local cost function is defined by the expression
\begin{equation}
\label{eq:local_cost_function}
    C_L(\theta) 
    = \frac{\hat{C_L}(\theta)}{\braket{\psi}{\psi}},
\end{equation}
where $\hat{C}_L(\theta) = \bra{x(\theta)}H_L\ket{x(\theta)}$ and $H_L = A^\dag U (\mathds{1} - \frac{1}{n} \sum_{q=0}^{n-1} \ket{0_q}\bra{0_q} \otimes \mathds{1}_{\bar{q}}) U^\dag A$. Here, $\mathds{1}$ denotes the $N$-dimensional identity matrix, and $\ket{0_q}\bra{0_q} \otimes \mathds{1}_{\bar{q}}$ represents the tensor product of the matrix $\ket{0}\bra{0}$ with $n-1$ identity matrices of dimension 2, where the matrix $\ket{0} \bra{0}$ is positioned at the $q$-th place.
The term ``local'' emphasizes that this cost function is computed with the summation of terms that consider information only from specific portions of the circuit.

In particular, considering the identity
\begin{equation}
    \label{eq:local_equivalence}
    \frac{1}{n} \sum_{q=0}^{n-1} \ket{0_q}\bra{0_q} \otimes \mathds{1}_{\bar{q}}
    = \frac{1}{2} \pars{\mathds{1} + \frac{1}{n} \sum_{q=0}^{n-1}Z_q },
\end{equation}
where $Z_q$ denotes the Pauli-Z operator locally applied to the $q$-th qubit, (\ref{eq:local_cost_function}) can be reformulated as
\begin{equation}
    \label{eq:local_cost_bis}
    C_L(\theta) = \frac{1}{2} - \frac{1}{2n} \frac{ \sum_{q=0}^{n-1} \bra{0} V^\dag A^\dag U Z_q U^\dag A V \ket{0}}{\braket{\psi}{\psi}}.
\end{equation}

Moreover, when considering the Pauli LCU decomposition, and assuming that the matrix $A$ is Hermitian, each term within the sum in the numerator of (\ref{eq:local_cost_bis}) can be written as
\begin{equation}
    \label{eq:real_loc_num}
    \begin{split}
        \bra{0} V^\dag A^\dag &U Z_q U^\dag A V \ket{0} 
        = \sum_{i=0}^{L-1} c_i^2 \bra{0} V^\dag A_i^\dag U Z_q U^\dag A_i V \ket{0} \\
        &+ \sum_{i=0}^{L-1} \sum_{j=i+1}^{L-1} 2 c_i c_j \Re{\bra{0} V^\dag A_i^\dag U Z_q U^\dag A_j V \ket{0}}.
    \end{split}
\end{equation}

Again, we note that the number of Hadamard tests for estimating the numerator required in the literature is $2L^2$. However, by using (\ref{eq:real_loc_num}), we only need $\frac{L(L-1)}{2} + 2L$ Hadamard tests. Therefore, if we adopt the LCU decomposition in the most general case, we previously needed to perform $2N^4$ Hadamard tests for the numerator computation, while now we only need $\frac{N^4 + 3N^2}{2}$ tests. For further details, we refer to Appendix \ref{cost_function_appendix} and \ref{circuit_execution_appendix}.

Local cost functions offer a notable advantage as the system scales, since they can mitigate the manifestation of barren plateaus compared to global cost functions. Specifically, previous research \cite{cerezo_2021_bp_local} has shown that in scenarios where the ansatz is an alternating layered ansatz composed of blocks that form local 2-designs (like the ansatz in Fig. \ref{fig:ansatz} used in our experiments), local cost functions exhibit gradients that scale at worst polynomially with the system size. This is in contrast with the exponential scaling observed with their global counterparts.
However, it is essential to acknowledge that the computation of the VQLS local cost function requires a higher number of circuit executions, approximately $n$ times those needed for the global cost function, where $n$ represents the number of qubits. This results in a non-negligible increase in the computational cost, which becomes $O(nL^2)$ instead of $O(L^2)$.

Similarly to the selection of the ansatz, the choice of the cost function is crucial for achieving efficiency in solving optimization problems when using a variational approach. Several studies have explored various cost functions with the goal of enhancing the algorithm effectiveness when addressing specific problems. Notably, \cite{sato_2021} addresses the Poisson problem with a variational approach using the Dirichlet energy as cost function. Furthermore, \cite{ali_2023} compares the efficacy of this cost function with the global and the local cost functions of the VQLS algorithm for the Poisson problem, showing that the Dirichlet energy yields better solutions with a smaller number of time steps.

\subsection{Previous Works using VQLS}

Several studies have explored the potential of VQLS in diverse applications. In the realm of linear equation systems arising from Partial Differential Equations (PDEs), Liu et al.\cite{liu_2024} addressed linear systems derived from the finite difference method applied to the Stokes flow problem, whereas Trahan et al. \cite{trahan_2023} tackled linear systems originated from the finite element discretization of the Poisson, heat, and wave equations.
Liu et al.\cite{liu_2022} applied VQLS to systems coming from the heat conduction problems and Shang et al.\cite{shang_2023} addressed linear systems with the goal of efficiently simulating the ocean circulation, incorporating also an analysis of the Zero Noise Extrapolation (ZNE) error mitigation technique to solutions obtained in real-hardware noisy scenarios.
Beyond these, VQLS finds application in diverse areas. Luo et al. \cite{luo_2024} applied VQLS to the Dempster–Shafer Theory (DST), whereas Xing et al. \cite{xing_2023} employed VQLS for multichannel quantum scattering problems, specifically in its step of matrix inversion.

Moreover, some authors have proposed variations to VQLS in order to enhance its efficacy. Patil et al. \cite{patil_2022} introduced a dynamic ansatz variant, in which new layers of gates are added to the ansatz during the iterations of the algorithm until a specific tolerance is met.
In \cite{huang_2021}, the authors observe that traditionally adopted ansatzes, such as agnostic hardware-efficient ansatzes (e.g., the one used in \cite{bravo-prieto_2023}) and alternating operator ansatzes \cite{hadfield_2019} (which depend on the specific linear system being solved), suffer from the barren plateau phenomenon, creating challenges in the search for the minimum of the cost function. To address this, they introduce an approach called the \textit{classical combination of variational quantum states} (CQS), in which the solution $x$ of the given linear system is expressed by the linear combination of variational quantum states, each created on the hardware: $x = \sum_i \alpha_i \ket{\psi_{V_i}(\vartheta_i)}$. It is worth noting that $x$ is never physically created on the quantum hardware and may not be normalized. Instead, only the individual states $\ket{\psi_{V_i}(\vartheta_i)}$ are produced. Although this approach requires executing more circuits to evaluate the cost function, it shows promise in mitigating the barren plateau phenomenon.
Yi et al. \cite{yi_2023} proposed VQLS-enhanced Quantum Support Vector Machine (QSVM), utilizing VQLS for solving linear equations associated with Least Squares Support Vector Machine (LS-SVM).
Pellow-Jarman et al. \cite{pellow-jarman_2023} integrated well-established techniques for variational algorithms with VQLS, introducing and evaluating multiple variants of the algorithm across diverse scenarios. The first introduced variant, the Adiabatic-Assisted Variational Quantum Linear Solver (AAVQLS), is a method inspired from adiabatic evolution, where the linear system matrix is varied over time. Another variant is the Evolutionary Ansatz VQLS (EAVQLS), which explores multiple ansatzes which are modified based on a genetic strategy during the algorithm execution. Lastly, the Logical Ansatz VQLS (LAVQLS), is a method that employs a linear combination of multiple ansatzes, offering a more flexible approach for solving the linear system.
Saito et al. \cite{saito_2023} proposed the Iterative Refinement VQLS (IR-VQLS), a variant of the algorithm which allows to obtain solutions with the same accuracy but requiring fewer measurements.

Finally, Pellow-Jarman et al. \cite{pellow-jarman_2021} conducted a comprehensive study on classical optimizers for fine-tuning parameters in VQLS circuits across noise-free, shot noise, and simulated hardware noise scenarios. The optimizer SPSA demonstrated superior efficacy in noisy scenarios, whereas COBYLA consistently exhibited the fastest convergence, reaching good-quality solutions in noise-free settings.

These studies provide valuable insights, but they mainly focus on small instances or system matrices with specific structures.
Since the VQLS algorithm was presented as applicable in a broader contest, it is important to investigate its performance on larger and more heterogeneous instances.
Additionally, the previous studies often do not analyze some issues like the convergence of the method and the high number of Hadamard tests required as the system scales. Our objective is to fill these gaps and evaluate the efficiency of VQLS, highlighting both its strengths and limitations.

\section{Experimental Analysis} \label{sec:experimental_protocol}
Since our goal is to investigate the effective applicability of VQLS to a heterogeneous class of problems with different sizes, we define various problem instances, apply the VQLS algorithm, and analyze the results.
This section details the problem instances chosen for testing the algorithm, as well as the specific decisions made in configuring the algorithm. These choices include the structure of the ansatz, the type of LCU decomposition, and the optimizer.

\subsection{Problem Instances}
\label{subsec:problem_instance}
We evaluate VQLS in three different scenarios. First, we replicate the same problem reported in the paper that originally proposed VQLS. Second, in order to assess the impact of the state preparation step we introduce more general instances which exhibit states that are difficult to prepare. Lastly, we consider a problem with real-world applications, specifically within the domain of fluid dynamics.

\paragraph{Ising}
We begin our analysis by addressing the same problem presented in the original paper by Bravo-Prieto et al. \cite{bravo-prieto_2023}, with the goal of ensuring we are able to obtain high-quality results on the same problem tested in the original article. This problem considers a linear system where the matrix $A$ is represented by the Ising Hamiltonian
\begin{equation}
\label{eq:ising}
    A = \frac{1}{\xi} \left(\sum_{i=1}^{N} X_i + J \sum_{i=1}^{N} Z_i Z_{i+1} + \eta \mathds{1} \right),
\end{equation}
and the vector \ket{b} is an equal superposition of states, obtained by applying a layer of Hadamard gates $H$ to the zero state:
\begin{equation}
    \ket{b} = H^{\otimes n} \ket{0}.
\end{equation}
This particular problem is referred to as the \textit{ising} Quantum Linear Systems Problem (QLSP). To ensure consistency, the parameters $J$, $\eta$, and $\xi$ are chosen to satisfy the conditions outlined in the paper. Specifically, $J = 0.1$, whereas $\eta$ and $\xi$ are chosen so that the maximum eigenvalue of $A$ is equal to 1. Given the presence of a free condition in the paper, we arbitrarily fix $\eta$ = 5.
For our experimental setup, we consider problem instances having $2, 4, 6, 8,$ and $10$ qubits.
The system matrices obtained have condition numbers with limited growth, whose specific values are 2.34, 9.08, 13.62, 20.16, and 30.38.

\paragraph{Random Pauli}
Then, we aim to evaluate the VQLS algorithm on linear systems characterized by more general matrices, and where the preparation of the quantum state \ket{b} is not trivial.
We define the matrices $A$ through their LCU decomposition (\ref{eq:LCU_decomposition}). To ensure consistency with the \textit{ising} QLSP, we set the number $L$ of unitary matrices to be $2n$ (where $n$ represents the number of qubits).
The matrices $A_i$ are tensor products of single Pauli matrices, uniformly chosen from the set $\{I, X, Z\}$. The vector $b$ comprises real numbers uniformly selected from the range $[-1, 1]$. The coefficients $c_i$ are real random values within the interval $[-10, 10]$, scaled to ensure that the resulting matrix $A$ has a maximum eigenvalue equal to 1.
This particular problem is denoted as the \textit{random pauli} problem.
Again, the considered instances involve a varying number of qubits, specifically $2, 4, 6, 8,$ and $10$ qubits.
The matrices generated with this procedure have condition numbers which grow exponentially in the number of qubits. Specifically, the condition numbers assume the values of 2.31, 3.20, 9.59, 72.70, 219.21. 

\paragraph{Darcy}
Finally, we extend our analysis to linear system problems derived from the broader and more realistic domain of fluid dynamics, which has already been done in \cite{trahan_2023, liu_2024, liu_2022}.
In this scenario, the matrix $A$ and the vector $b$ of the linear system are derived using the finite elements method applied to the equations modeling fluid flow in porous media. 
In this problem, the geometry of the physical system varies over time due to compaction, possible erosion of the sedimented material, and sedimentation of new materials.
The system consists of two phases: the solid rocks and the fluid, both of which are in motion.
We are often interested in computing the overpressure of the liquid phase inside the pores. The overpressure is defined as $u(x,y,z,t) = p(x,y,z,t) - p_h(z,t)$, where $p(x,y,x,t)$ is the pore pressure (i.e. the pressure of the liquid inside the pores) in the point $(x,y,z)$ at time $t$, and $p_h(z,t)$ is the hydrostatic pressure. For the case of stationary flow, the overpressure gradient is the driving force for the fluid flow. This relation is modelled through the Darcy's equation:
\begin{equation}
    \phi \left(\vec{v} -\vec{v}_S\right)= - \frac{\mathbf{K}}{\mu} \nabla u.
\end{equation}
Here, $\phi$ is the porosity of the rocks, $\vec{v}$ represents the fluid velocity, $\vec{v}_S$ is the velocity of the solid phase, $\mu$ is the fluid viscosity, and $\mathbf{K}$ is the permeability of the medium. In general, for non-isotropic medium, $\mathbf{K}$ is a tensor.
By considering the mass conservation equations of the fluid and the solid phase, we obtain the equation modelling the fluid flow in a porous medium subjected to compression\cite{hantschel_2009}:
\begin{equation}
- \nabla \cdot \frac{\mathbf{K}}{\mu} \cdot \nabla u = \frac{C}{1-\phi}\frac{\partial}{\partial t} (u-u_l),
\end{equation}
where $u_l$ is defined as the difference between the lithostatic pressure and the hydrostatic pressure ($u_l = p_l - p_h$), and $C$ is the compression factor modelling variation in the porosity due to compaction. The compression factor $C$ is determined in order to satisfy the Terzaghi's law\cite{hantschel_2009}:
\begin{equation}
    \frac{\partial \phi}{\partial t} = -C \frac{\partial \sigma'_z}{\partial t},
\end{equation}
where $\sigma'_z$ is the vertical component of the effective stress $\sigma'$, defined as $\sigma'=\sigma - p I$, here $\sigma$ is the stress, $p$ is the fluid pressure, and $I$ the identity matrix.
We denote this problem as \textit{darcy} problem.

In this domain, the system matrices obtained through the finite element method are symmetric and exhibit a distinct block structure. Specifically, the lower-right and upper-left blocks are equal and contain only null components.
Moreover, since the linear system is obtained after having imposed the boundary conditions on the boundary nodes, the lower-left block is diagonal.
The remaining upper-right block is a sparse matrix which presents strict bands of non-null elements. Since the solution components corresponding to the diagonal block can be computed trivially and independently from the others, we focus only on this sparse block.
To facilitate efficient computation, the sparse block matrix and the corresponding portion of the vector $b$ undergo a normalization process. Each element is divided by the maximum eigenvalue of the sparse block, resulting in a sparse matrix with a maximum eigenvalue of 1.
However, it is important to note that the size of such matrix may not necessarily be a power of 2, a requirement for employing VQLS algorithm. To overcome this limitation, we adopt a padding strategy by adding columns and rows to achieve a total size equal to the closest power of 2 greater than the original size. The added rows and columns are filled with zeros, except for the diagonal components, which are set to ones. Simultaneously, the size of the $b$ vector is adjusted by appending zeros to match the new number of columns (and rows) in the matrix. As a result, the solution components added with this padding strategy will be equal to 0.
To be noted, then, that the initial elimination of the lower-left diagonal block is convenient only if the smaller integer grater than or equal to the $log_2$ of the size of the matrix is reduced.
We address problem instances involving sparse matrices $A$ with size of $20 \times 20$ and $1140 \times 1140$. The upper-right blocks  have sizes $12 \times 12$ and $656 \times 656$ respectively. Consequently, following the aforementioned modifications, we obtain problem instances requiring 4 and 10 qubits respectively.
These matrices have a condition numbers equal to 400793.62 and 19051.15 respectively. The $12 \times 12$ matrix is actually a dense matrix due to the fact that we wanted to address the smallest possible problem. 

\subsection{VQLS configuration}

In this section we detail the specific choices made for configuring the VQLS algorithm, including ansatz, cost function, and classical optimizer.

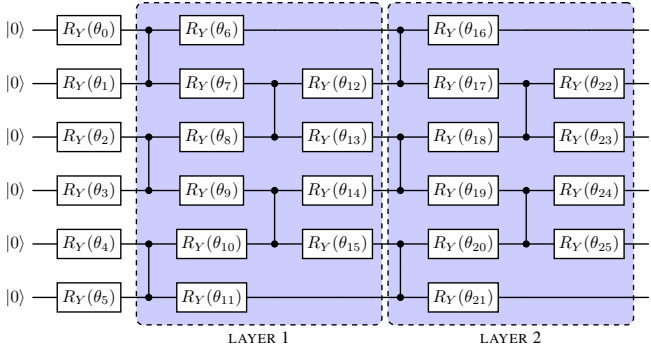
\begin{figure}
    \centering
    \resizebox{\columnwidth}{!}{
    \begin{quantikz}[wire types={q,q,q,q,q,q}]
        \lstick{\ket{0}} & \gate{R_Y(\theta_0)} & \ctrl{1}\gategroup[6,steps=4,style={dashed, rounded corners,fill=blue!20, inner xsep=2pt},background,label style={label position=below,anchor=north,yshift=-0.2cm}]{{\sc layer 1}} & \gate{R_Y(\theta_6)} &&& \ctrl{1}\gategroup[6,steps=4,style={dashed, rounded corners,fill=blue!20, inner xsep=2pt},background,label style={label position=below,anchor=north,yshift=-0.2cm}]{{\sc layer 2}} & \gate{R_Y(\theta_{16})} &&& \\
        \lstick{\ket{0}} & \gate{R_Y(\theta_1)} & \control{} & \gate{R_Y(\theta_7)} & \ctrl{1} & \gate{R_Y(\theta_{12})} & \control{} & \gate{R_Y(\theta_{17})} & \ctrl{1} & \gate{R_Y(\theta_{22})} & \\
        \lstick{\ket{0}} & \gate{R_Y(\theta_2)} & \ctrl{1} & \gate{R_Y(\theta_8)} & \control{} & \gate{R_Y(\theta_{13})} & \ctrl{1} & \gate{R_Y(\theta_{18})} & \control{} & \gate{R_Y(\theta_{23})} & \\
        \lstick{\ket{0}} & \gate{R_Y(\theta_3)} & \control{} & \gate{R_Y(\theta_9)} & \ctrl{1} & \gate{R_Y(\theta_{14})} & \control{} & \gate{R_Y(\theta_{19})} & \ctrl{1} & \gate{R_Y(\theta_{24})} & \\
        \lstick{\ket{0}} & \gate{R_Y(\theta_4)} & \ctrl{1} & \gate{R_Y(\theta_{10})} & \control{} & \gate{R_Y(\theta_{15})} & \ctrl{1} & \gate{R_Y(\theta_{20})} & \control{} & \gate{R_Y(\theta_{25})} & \\
        \lstick{\ket{0}} & \gate{R_Y(\theta_5)} & \control{} & \gate{R_Y(\theta_{11})} &&& \control{} & \gate{R_Y(\theta_{21})} &&&
    \end{quantikz}
    }
    \caption{Ansatz employed in our experiments. It consists of initial single-qubit $R_Y$ rotation gates, followed by multiple layers whose number is controlled by the hyperparameter \textit{layers}. Each layer is composed of $CZ$ gates acting on alternating pairs of neighboring qubits, followed by $R_Y$ rotation gates on each qubit. This is followed by another set of $CZ$ gates on alternating pairs of neighboring qubits, starting from the second qubit, and a final set of $R_Y$ gates.\label{fig:ansatz}}
\end{figure}

\paragraph{Ansatz}
To assess the effectiveness of the VQLS algorithm, we adopt the same ansatz described in the original paper by Bravo-Prieto et al. \cite{bravo-prieto_2023} (see Fig. \ref{fig:ansatz}). This ansatz employs only $R_Y$ and $CZ$ gates, ensuring that the final quantum state has real amplitudes. The depth of the circuit can be adjusted by varying the hyperparameter \textit{layers}, which expands the solution space as \textit{layers} increases. The circuit parameters are the angles of the $R_Y$ rotation gates.
We evaluate the VQLS algorithm on each problem instance using this ansatz varying the values of the hyperparameter \textit{layers} from 1 to 5.

\paragraph{Cost Function}

Our goal is to study the effectiveness of VQLS on more heterogeneous and larger problems. Since we use a simulation of the algorithm, the estimation of the cost function value can be obtained without performing the LCU decomposition and the execution of the Hadamard tests for computing the terms in (\ref{eq:real_glob_den}), (\ref{eq:real_glob_num}), and (\ref{eq:real_loc_num}). We consider matrices representing the ansatz and the Hamiltonian in (\ref{eq:global_cost_function}) and (\ref{eq:local_cost_bis}) and directly compute the matrix products. The two procedures are mathematically equivalent, but the latter is more scalable and eliminates potential approximations introduced by the Hadamard test.

\paragraph{Optimizer}
The circuit parameters are optimized by adopting the COBYLA optimizer\cite{powell_1994_cobyla}\footnote{We use the implementation provided in the SciPy library: \url{https://qiskit-community.github.io/qiskit-algorithms/stubs/qiskit_algorithms.optimizers.COBYLA.html}.}, which is largely used in quantum variational approaches due to its effectiveness in noise-free scenarios without demanding an excessive computational cost\cite{singh_2023_cobyla2, fernandez-pendas_2020_cobyla3}. Moreover, in a dedicated comparison of optimizers within the context of VQLS\cite{pellow-jarman_2021}, COBYLA demonstrated the fastest convergence while ensuring good-quality results.
The initial parameters for optimization are randomly selected, so that the resulting rotation angles are uniformly distributed in $[0,2\pi)$.

The algorithm terminates its execution when the minimum cost function value found does not improve for 100 consecutive iterations of the optimizer. Additionally, to ensure the experiments are allocated reasonable computational resources we set a maximum number of 100,000 iterations. Upon meeting the termination condition, the algorithm stops, and the circuit parameters associated with the lowest cost value encountered during the optimization process are considered as the final result.

\subsection{Evaluation Metrics}

The evaluation metrics include the cosine of the angle between $Ax$ and $b$, along with the final value of the cost function. Given that our primary objective is to find a solution $x$ such that $Ax$ is proportional to $b$\footnote{The VQLS algorithm is designed to find a solution $x$ with unitary norm. To obtain a solution $\tilde{x}$ with the correct norm, one can apply the normalization $\tilde{x}=x\frac{\norm{b}}{\norm{Ax}}$.}, achieving a cosine having absolute value close to 1 is our main goal. It is worth noting that if the identified solution $x$ results in $Ax$ being parallel to $b$, the cost function attains a value of 0. Consequently, cost function values approaching 0 are desirable.

We perform ten executions of the algorithm with different random initial parameters for each problem instance and then average the results.

\section{Results}

In this section, we present the results of the application of VQLS algorithm on the problem instances described in Section \ref{sec:experimental_protocol} under the specified conditions.

\subsection{VQLS effectiveness}
\label{sec:results_quality}

Table \ref{tab:reps_optimization} shows the average results of the experiments described in Section \ref{sec:experimental_protocol}, across 10 executions of VQLS for each problem instance.
Each row corresponds to a specific problem instance and indicates the number of layers in the ansatz that yielded the highest average absolute value of the cosine of the angle between $Ax$ and $b$, truncated to three decimal places, both for the global and the local cost function. When ansatzes with different numbers of layers resulted in the same cosine absolute value, the one with the smallest number of layers was chosen.
Furthermore, we report the average cosine absolute value, the average minimum cost achieved, and the average number of cost function evaluations associated with the ansatz having the optimal number of layers, along with their standard deviations.

Our analysis reveals that the algorithm effectiveness is closely tied to the specific problem being addressed. For the \textit{ising} problem, the algorithm consistently reaches a cosine of the angle with an absolute value near 1, indicating very accurate results. Consistently, the minimum cost achieved remains close to 0.
Additionally, when shifting focus to the \textit{random pauli} problem, the cosine of the angle in the instances with $n=8$ and $n=10$ qubits is respectively below 0.75 and below 0.50 for both the global and the local cost function.
A similar trend emerges in the \textit{darcy} problem, where the cosine values are consistently below 0.83.
Moreover, the low standard deviations indicate that these sub-optimal results are consistently obtained across multiple executions of the algorithm.
The inability of VQLS to reach cosines close to 1 can be attributed to two primary reasons: either the optimizer is inefficient in finding the global optimum or the chosen ansatz does not contain the solution of the problem within its space of solutions. We delve into these aspects in Section \ref{sec:sub-optimal_results}.

We can also note that the optimal number of layers varies significantly depending on the specific problem being addressed.
For the \textit{ising} problem, the optimal value of \textit{layers} is consistently 1. This observation suggests that even a single layer in the ansatz is sufficient for obtaining high-quality solutions for this particular problem.
Further investigation into the reasons for these good results is reported in \ref{sec:results_quality}.
In contrast, for the \textit{random pauli} problem, the best \textit{layers} value is equal to 1 only for the problem instance with $n=2$, increasing to a value of 4 for $n=4$, and reaching the maximum tested value of 5 for all the other instances. Here, it is evident that using circuits with more than one layer is essential to improve the solution quality. 
In particular, for the instances with $n=8$ and $n=10$, a number of layers beyond 5 would likely contribute to further improve the results quality. However, deep circuits are impractical for real-world applications, as they result in long computational times and are more prone to noise. Therefore, we decide not to analyze the algorithm performance with a larger number of layers in the ansatz.
For the \textit{darcy} problems, the algorithm continues to encounter challenges in reaching a cost function value of 0. Interestingly, the optimal number of layers for the instance with $n=4$ is 1. The lack of improvement with an increasing number of layers is likely attributable to the higher difficulty the optimizer faces in finding the global optimum as the number of layers - and consequently, the number of parameters - increases, or it may be due to the inadequacy of the ansatz for this particular problem instance.

\begin{table*}[ht]
\centering
\caption{Results with the best number of layers. For each problem instance the optimal value for \textit{layers}, the final cost function value, the absolute value of the cosine of the angle between $Ax$ and $b$, and the number of cost function evaluations are reported. The results are presented as the average and the standard deviation across 10 executions of VQLS.\label{tab:reps_optimization}} 
\resizebox{\linewidth}{!}{
\begin{tabular}{cc|cccc|cccc}
\toprule
\multicolumn{2}{c|}{} & \multicolumn{4}{c|}{\textbf{Global Cost Function}} & \multicolumn{4}{c}{\textbf{Local Cost Function}}\\
\textbf{problem} & \textbf{n} & \textbf{\begin{tabular}{@{}c@{}}best\\layers\end{tabular}} & \textbf{min cost} & \textbf{cos Ax b} & \textbf{iterations} & \textbf{\begin{tabular}{@{}c@{}}best\\layers\end{tabular}} & \textbf{min cost} & \textbf{cos Ax b} & \textbf{iterations}\\ 
\midrule 
\multirow{5}{*}{ising}
    & 2 & 1 & $\sim$ 0 & 0.999 ± 0\phantom{.000} & \phantom{000}415 ± 210\phantom{00} & 1 & $\sim$ 0 & 0.999 ± 0\phantom{.000} & \phantom{000}362 ± 112\phantom{00} \\ 
    & 4 & 1 & $\sim$ 0 & 0.999 ± 0\phantom{.000} & 100000 ± 0\phantom{0000} & 1 & $\sim$ 0 & 0.999 ± 0\phantom{.000} & \phantom{0}88310 ± 12907 \\ 
    & 6 & 1 & $\sim$ 0 & 0.999 ± 0\phantom{.000} & 100000 ± 0\phantom{0000} & 1 & $\sim$ 0 & 0.999 ± 0\phantom{.000} & 100000 ± 0\phantom{0000} \\ 
    & 8 & 1 & $\sim$ 0 & 0.999 ± 0\phantom{.000} & 100000 ± 0\phantom{0000} & 1 & $\sim$ 0 & 0.999 ± 0\phantom{.000} & 100000 ± 0\phantom{0000} \\ 
    & 10 & 1 & $\sim$ 0 & 0.999 ± 0\phantom{.000} & 100000 ± 0\phantom{0000} & 1 & $\sim$ 0 & 0.999 ± 0\phantom{.000} & 100000 ± 0\phantom{0000} \\ 
\midrule 
\multirow{5}{*}{\begin{tabular}{c@{}}random \\ pauli\end{tabular}}
    & 2 & 1 & $\sim$ 0 & 0.999 ± 0\phantom{.000} & \phantom{000}772 ± 372\phantom{00} & 1 & $\sim$ 0 & 0.999 ± 0\phantom{.000} & \phantom{000}704 ± 283\phantom{00} \\ 
    & 4 & 4 & $\sim$ 0 & 0.999 ± 0\phantom{.000} & \phantom{0}12993 ± 13184 & 4 & $\sim$ 0 & 0.999 ± 0\phantom{.000} & \phantom{0}18651 ± 16556 \\ 
    & 6 & 5 & 0.028 ± 0.012 & 0.985 ± 0.006 & 100000 ± 0\phantom{0000} & 5 & 0.014 ± 0.007 & 0.976 ± 0.013 & 100000 ± 0\phantom{0000} \\ 
    & 8 & 5 & 0.436 ± 0.017 & 0.750 ± 0.011 & 100000 ± 0\phantom{0000} & 5 & 0.191 ± 0.010 & 0.637 ± 0.038 & 100000 ± 0\phantom{0000} \\ 
    & 10 & 5 & 0.755 ± 0.008 & 0.493 ± 0.008 & 100000 ± 0\phantom{0000} & 5 & 0.341 ± 0.006 & 0.255 ± 0.044 & 100000 ± 0\phantom{0000} \\ 
\midrule 
\multirow{2}{*}{darcy}
    & 4 & 1 & 0.402 ± 0\phantom{.000} & 0.772 ± 0\phantom{.000} & \phantom{0}31881 ± 21767 & 1 & 0.107 ± 0\phantom{.000} & 0.772 ± 0\phantom{.000} & \phantom{0}94262 ± 12421 \\ 
    & 10 & 5 & 0.316 ± 0.050 & 0.825 ± 0.030 & 100000 ± 0\phantom{0000} & 5 & 0.117 ± 0.022 & 0.651 ± 0.123 & 100000 ± 0\phantom{0000} \\ 
\bottomrule 
\end{tabular}
}
\end{table*}

\paragraph{Varying $J$ in the \textit{ising} problem instances}
\begin{table}[ht]
\centering
\caption{Results obtained for the \textit{ising} problem with $n=8$ qubits described in (\ref{eq:ising}), with $J=5$. The table presents the cosine values, final cost values, and the number of evaluations for different numbers of layers in the ansatz.\label{tab:ising_J}} 
\begin{tabular}{cccc}
    \toprule
        \textbf{layers} & \textbf{min cost} & \textbf{cos Ax b} & \textbf{iterations} \\ 
        \midrule
        1 & 0.456 ± 0.015 & 0.737 ± 0.010 & \phantom{0}56191 ± 43441 \\ 
        2 & 0.352 ± 0.009 & 0.805 ± 0.006 & \phantom{0}83440 ± 18744 \\ 
        3 & 0.264 ± 0.039 & 0.858 ± 0.023 & 100000 ± 0\phantom{0000} \\ 
        4 & 0.156 ± 0.034 & 0.919 ± 0.019 & 100000 ± 0\phantom{0000} \\ 
        5 & 0.109 ± 0.019 & 0.944 ± 0.010 & 100000 ± 0\phantom{0000} \\ 
    \bottomrule 
\end{tabular}
\end{table}

In this paragraph, we aim to investigate the factors that contribute to achieving high-quality results when solving the \textit{ising} problem.
In this analysis, we maintain the structure of the \textit{ising} problem as described in \ref{subsec:problem_instance}, but we set the value of the parameter $J$ in (\ref{eq:ising}) to 5, instead of the original 0.1. We choose a problem instance with $n=8$ qubits and execute the VQLS algorithm using the global cost function and the ansatz in Fig. \ref{fig:ansatz} varying the number of layers from 1 to 5. Each experiment is performed 10 times and the average results along with their standard deviations are reported in Table \ref{tab:ising_J}.

We can observe that the quality of results for the new \textit{ising} instances with $J=5$ is not as high as for the previous instances with $J=0.1$, both in terms of the cosine similarity between $Ax$ and $b$ and the final cost value reached.
This difference is likely due to the inadequacy of the employed ansatz in sampling the solution of the new problem.
Specifically, when $J=0$, the ansatz employed in Fig. \ref{fig:ansatz} contains the correct solution within the space of states that can be sampled. This is evident because \ket{b} of the \textit{ising} problem is an equal superposition of states and is an eigenvector of the matrix $A$, therefore is a solution to such problem. Notably, the ansatz in Fig. \ref{fig:ansatz} is capable of producing an equal superposition of states even with a single layer of gates, as $R_Y(\theta)$ rotations with an angle of $\frac{\pi}{2}$ are equivalent to Hadamard gates.
Therefore, when $J=0.1$, like in the original paper \cite{bravo-prieto_2023}, the perturbation in the cost function is limited, and the ansatz in Fig. \ref{fig:ansatz} still yields good results.
However, increasing $J$ to 5 introduces a more significant perturbation, resulting in lower cosine absolute values compared to the previous instances. Although the cosine increases with a higher number of layers, even with 5 layers it only reaches 0.93, differing from the nearly 1 observed previously.
It is worth noting that for more general $A$ matrices and $b$ vectors, selecting an appropriate ansatz capable of producing the solution to the linear system is not straightforward, and further investigation in this area is needed.

\subsection{Number of layers}

In this section we analyze the difference in the quality of the results varying the number of layers in the circuit. We consider six different problem instances: \textit{ising}, \textit{random pauli} and \textit{darcy}, each with $n=4$ and $n=10$ qubits, focusing on the global cost function. The results for the local cost function are omitted since they are analogous to those of the global cost function.

\begin{figure*}[htp]
    \centering
    \begin{subfigure}{0.49\textwidth}
        \includegraphics[width=\linewidth]{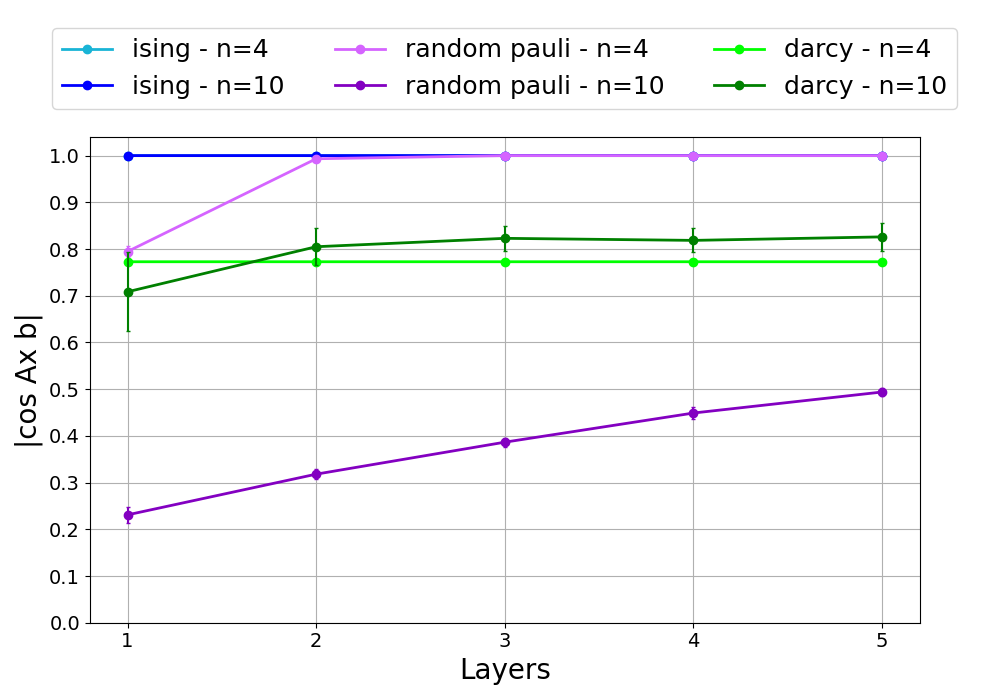}
        \caption{Absolute value of the cosine of the angle between $A\ket{x}$ and $\ket{b}$ obtained by optimizing the global cost function using ansatzes with a different number of layers across different problem instances. The reported values are averages over 10 executions of the algorithm, with error bars indicating the standard deviations.\label{fig:cosine_reps_global}}
    \end{subfigure}\hfill
    \begin{subfigure}{0.49\textwidth}
        \includegraphics[width=\linewidth]{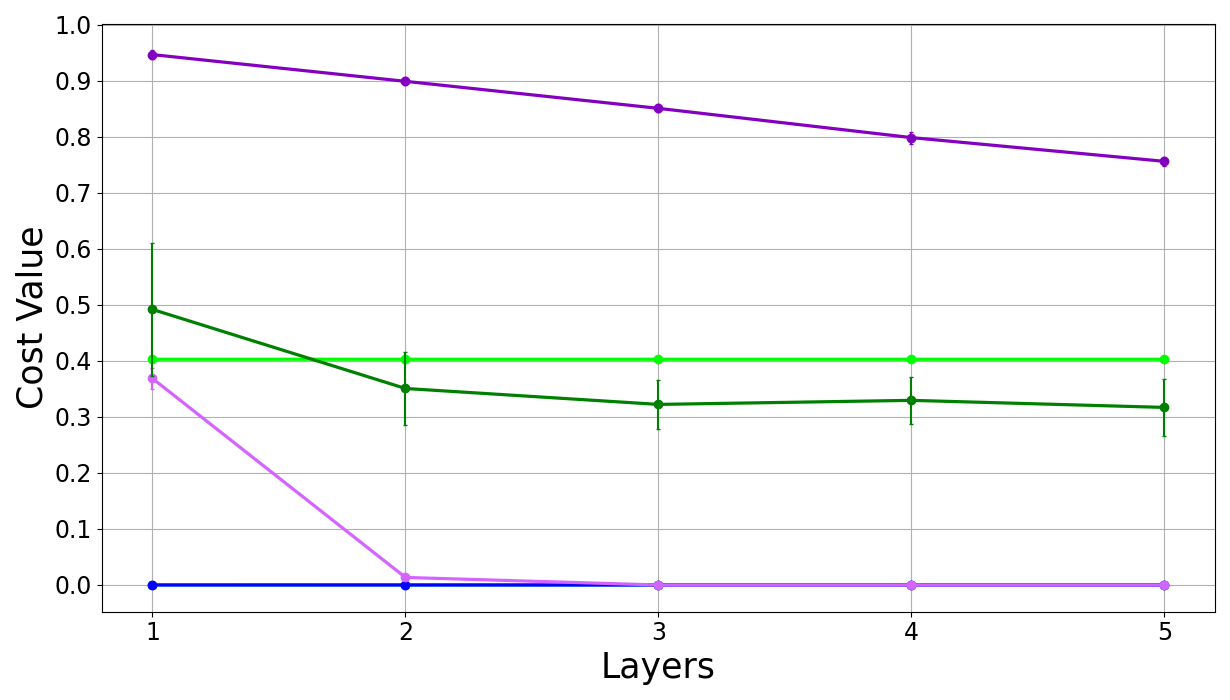}
        \caption{Final cost obtained by optimizing the global cost function using ansatzes with a different number of layers across different problem instances. The reported values are averages over 10 executions of the algorithm, with error bars indicating the standard deviations.\phantom{xxxxxxxxxxxxxxxxxxxxxxxxxxxxxxxx}\label{fig:cost_reps_global}}
    \end{subfigure}
    \caption{Cosine absolute value and final cost as functions of the number of layers. Note that the plots for the \textit{ising} instances are overlapping. \label{fig:cosine_and_cost}}
\end{figure*}

Fig. \ref{fig:cosine_reps_global} shows the absolute value of the cosine of the angle between $Ax$ and $b$ for the aforementioned problem instances, varying the number of layers. 
We observe different trends in the cosine depending on the specific problem considered.
For the \textit{ising} instances, the cosine is consistently around 1 for all tested numbers of layers, indicating that a low number of layers is sufficient for obtaining satisfactory results.
In contrast, for the \textit{random pauli} problem with $n=4$ qubits, 1 layer is insufficient, but increasing the number of layers to 2 results in a notable improvement, with the cosine approaching 1.
For the \textit{random pauli} problem with $n=10$, the results with 1 layer are still poor and improve as the number of layers increases, but, even with 5 layers, the results are not satisfactory.
The \textit{darcy} problem exhibits a different behavior. For the $n=4$ instances, 1 layer yields poor results and increasing the number of layers does not lead to any improvement. Similarly, for the $n=10$ instances, a non-negligible improvement is observed when increasing from 1 to 2 layers, but only marginal improvement is seen when increasing from 2 to 5 layers.

Similar trends can be observed in the graph representing the cost function value obtained at the end of the optimization process as a function of the number of layers. For all the \textit{ising} instances, the final cost is close to 0. In the case of the $n=4$ \textit{random pauli} instance, the final cost is far from zero with 1 layer but becomes closer to 0 as the number of layers increases.
However, for the $n=10$ \textit{random pauli} instance, the cost cannot reach 0 even with 5 layers.
Finally, for the \textit{darcy} instances with $n=4$, the final cost hovers around 0.4 for all \textit{layers} values, and it is never lower than 0.3 for the $n=10$ instances.

These results suggest that the dependence of the algorithm effectiveness on the number of layers is closely tied to the specific problem instance. In some cases, increasing the number of layers improves the quality of results, as observed in the \textit{random pauli} instances. However, in other cases, such as the \textit{darcy} problem instances, it has only marginal or almost no effect.
Therefore, we are led to believe that the potential advantage in using the VQLS should be searched within specific classes of problems. An open research question remains to understand which problems are best suited to benefit from the scaling advantages offered by VQLS.

\subsection{Cost function convergence}

\begin{figure*}[htp]
    \centering
    \begin{subfigure}{0.49\textwidth}
        \includegraphics[width=\linewidth]{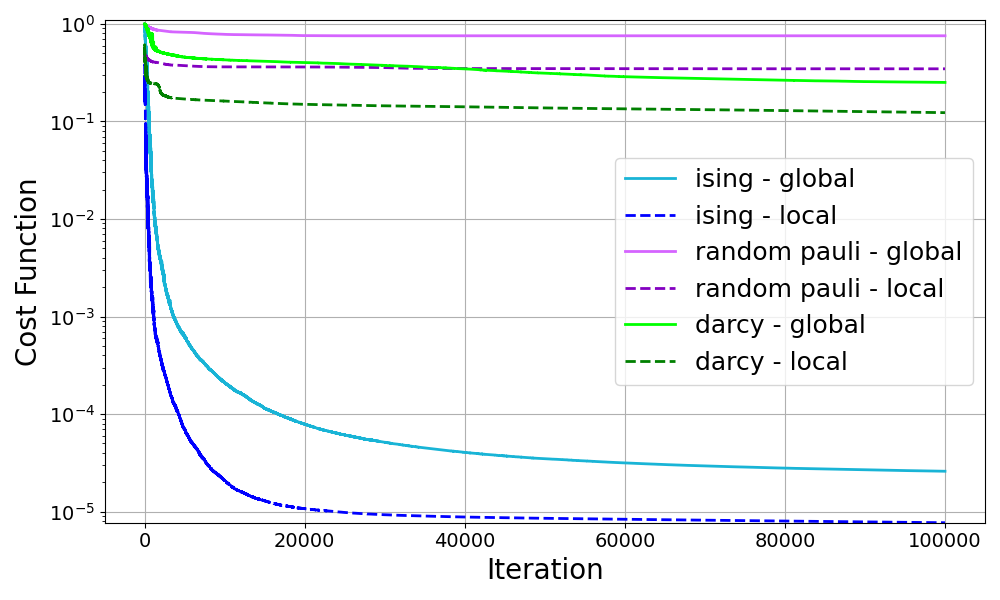}
        \caption{Global and local cost function convergence for \textit{ising}, \textit{random pauli}, and \textit{darcy} problem instances with $n=10$.\label{fig:convergence_global_local}}
    \end{subfigure}\hfill
    \begin{subfigure}{0.49\textwidth}
        \includegraphics[width=\linewidth]{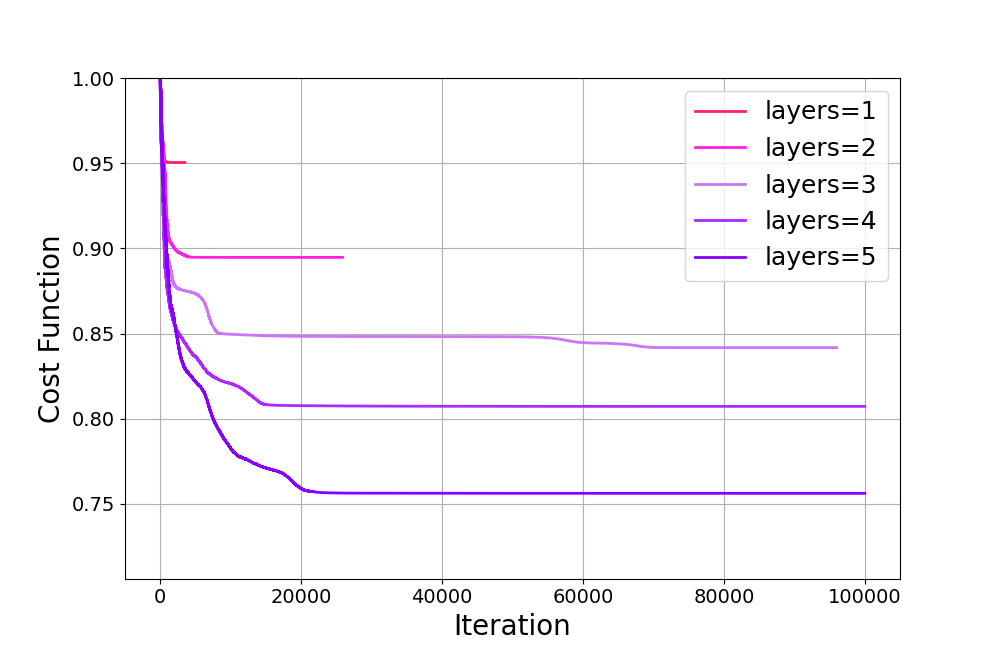}
        \caption{Global cost function convergence for the \textit{random pauli} problem instance with $n=10$ for different numbers of layers in the ansatz.\label{fig:convergence_different_reps_random_pauli}}
    \end{subfigure}
    \caption{Cost function convergence.\label{fig:cost_func_convergence}}
\end{figure*}

In this subsection we analyze the convergence of the cost function. Fig. \ref{fig:convergence_global_local} shows in a logarithmic scale the values of the global and the local cost functions attained across the iterations of a single algorithm execution, for three specific problem instances: \textit{ising}, \textit{random pauli}, and \textit{darcy}, all with $n=10$ qubits.
As shown by the figure, only the cost function related to the \textit{ising} problem converges to zero. Notably, the most significant reduction in the cost function value occurs within the initial 100 iterations, reaching a value below $10^{-2}$. This indicates that good results can be achieved even with a limited number of iterations.
For the remaining problem instances, the cost function appears to converge to a value with a non-negligible distance from zero. This behavior could be attributed either to the inadequacy of the ansatz in generating the state representing the solution to the linear system or to the challenges encountered by the classical optimizer in finding the global optimum.

Fig. \ref{fig:convergence_different_reps_random_pauli} illustrates the convergence of the cost function, presented in a linear scale, for a \textit{random pauli} problem instance with $n=10$, using ansatzes with a varying numbers of layers, ranging from 1 to 5.
Firstly, we can observe that the number of iterations actually executed by the algorithm is directly influenced by the number of layers: instances with fewer layers tend to reach convergence more quickly, as they involve fewer parameters to optimize. Secondly, as the number of layers increases, the final value of the cost function decreases.
It is important to highlight that the rate of reduction in the cost function value is rather irregular across the iterations, as evident in the instances with \textit{layers} = 3, 4, and 5. In particular, for the instance with \textit{layers} = 3, the cost function appears to plateau at around 10,000 iterations but continues to decrease non-negligibly after approximately 50,000 iterations.

\subsection{Circuit Resources}

A relatively unexplored aspect in the existing literature concerns the circuit resources required for estimating the cost function value. This consideration assumes particular importance when implementing the algorithm on real quantum devices, since excessively deep circuits could preclude the practical utilization of the hardware due to the increased noise and decoherence effects.

We aim to understand the depth of the individual circuits involved in each Hadamard test, necessary for estimating the real or imaginary parts of the expectations in the expressions of the global (\ref{eq:global_cost_function}) and local (\ref{eq:local_cost_function}) cost functions. However, we present only the results for the global cost function, as analogous considerations can be made for the local cost function.

\begin{figure*}[htp]
    \centering
    \begin{subfigure}{0.49\textwidth}
        \includegraphics[width=\linewidth]{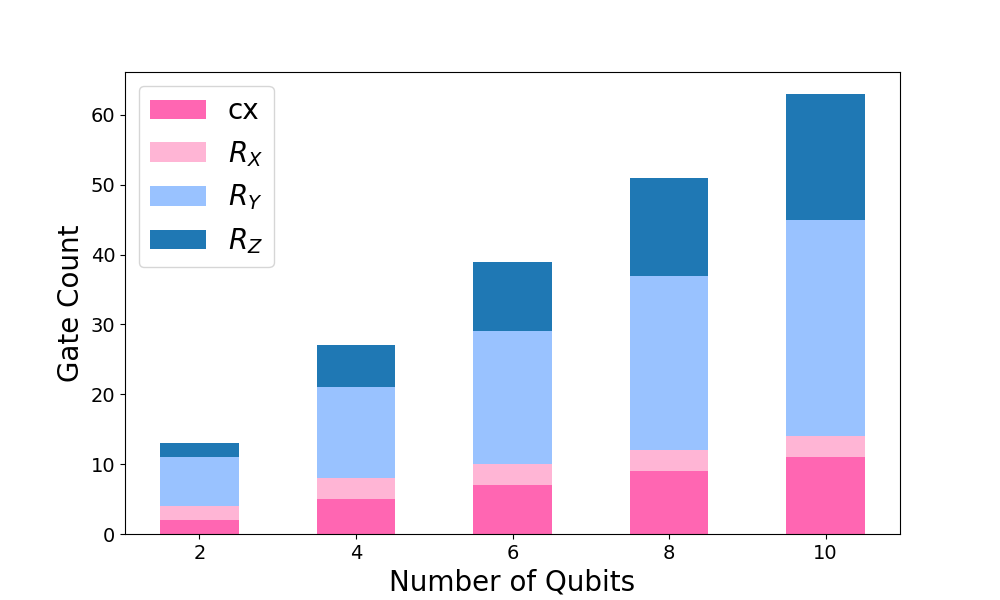}
        \caption{Gate count for the Hadamard test in the denominator of the global cost function for the \textit{ising} problem.\label{fig:gate_count_den_ising}}
    \end{subfigure}\hfill
    \begin{subfigure}{0.49\textwidth}
        \includegraphics[width=\linewidth]{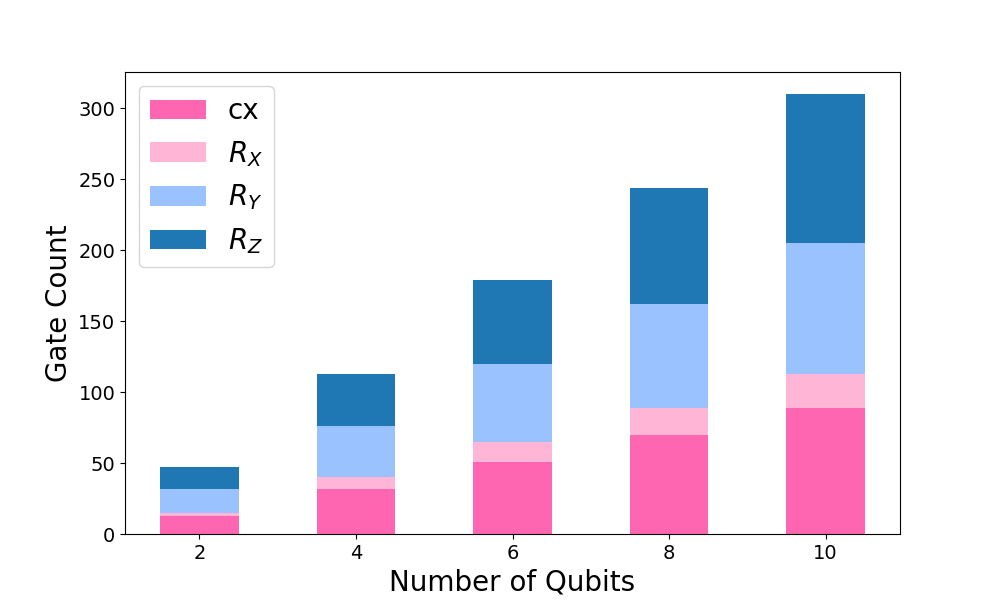}
        \caption{Gate count for the Hadamard test in the numerator of the global cost function for the \textit{ising} problem.\label{fig:gate_count_num_ising}}
    \end{subfigure}

    \medskip

    \begin{subfigure}{0.49\textwidth}
        \includegraphics[width=\linewidth]{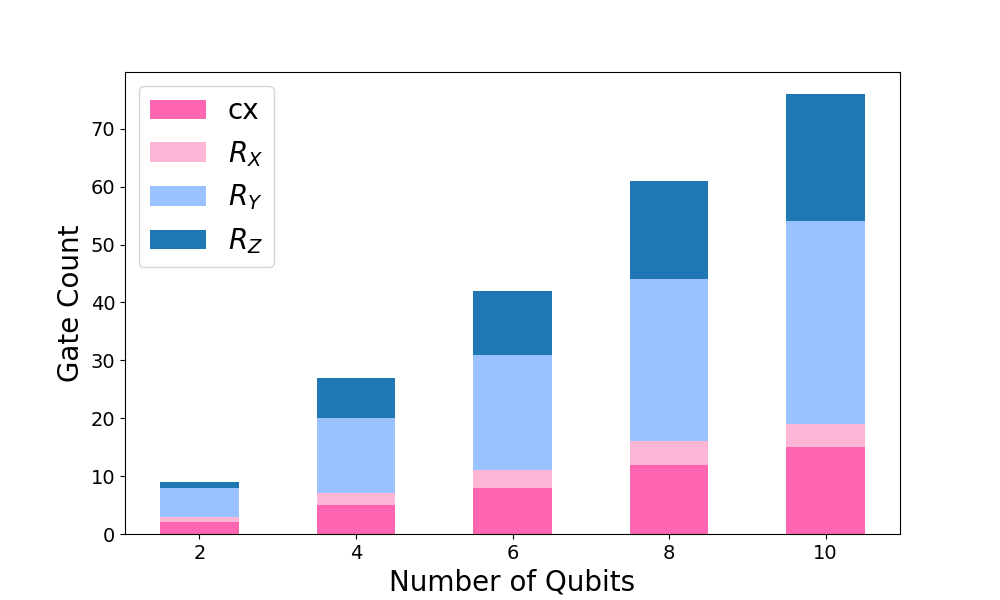}
        \caption{Gate count for the Hadamard test in the denominator of the global cost function for the \textit{random pauli} problem.\label{fig:gate_count_den_random_pauli}}
    \end{subfigure}\hfill
    \begin{subfigure}{0.49\textwidth}
        \includegraphics[width=\linewidth]{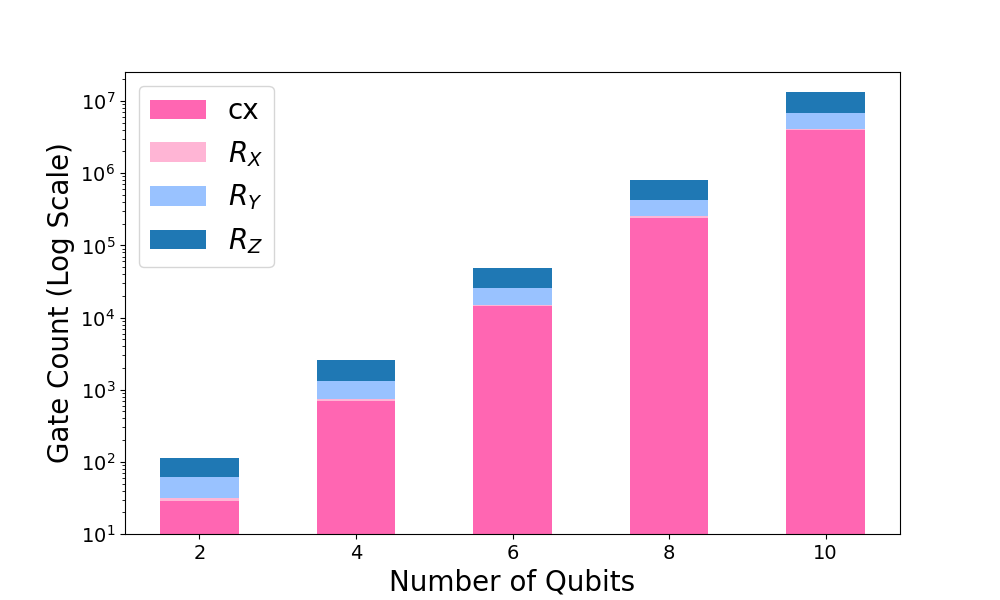}
        \caption{Gate count for the Hadamard test in the numerator of the global cost function for the \textit{random pauli} problem.\label{fig:gate_count_num_random_pauli}}
    \end{subfigure}
    \caption{Average gate count for Hadamard test circuits.\label{fig:gate_count}}
\end{figure*}

\paragraph{Gate count}
The histograms in Fig. \ref{fig:gate_count} show the counts of the gates for both \textit{ising} and \textit{random pauli} problem instances with a number of qubits set to $n=2, 4, 6, 8$, and $10$. The analysis focuses on the circuits of the Hadamard tests for computing the terms in the denominator (\ref{eq:real_glob_den}) and in the numerator (\ref{eq:real_glob_num}) of the global cost function.

Specifically, for each real or imaginary part of the expectations in (\ref{eq:real_glob_den}) and (\ref{eq:real_glob_num}), we build the Hadamard test circuits necessary to estimate that quantity. We then convert the circuit using only the gates $R_X$, $R_Y$, $R_Z$, and $CX$, employing the Qiskit transpiler with optimization level 3 (which is the highest possible)\footnote{The execution of the code for performing transpilation with optimization level 3 and gate count took around 40 days. We also performed the transpilation with the default optimization level 1 and this took around 20 days. We note that passing from optimization level 1 to 3 in the \textit{random pauli} instances led only to a small improvement in circuit depth and gate count.} and compute the gate count for each gate typology. Since the gate count depends on the specific circuit implemented (influenced by the specific matrices $A_i$), we calculate the average counts for the types of gates across all expectations in the denominator (Fig. \ref{fig:gate_count_den_ising}, \ref{fig:gate_count_den_random_pauli}) and numerator (Fig. \ref{fig:gate_count_num_ising}, \ref{fig:gate_count_num_random_pauli}).
It is important to note that each histogram provides information about the average resources for one single Hadamard test circuit, and that multiple Hadamard tests need to be run for estimating the numerator and denominator of the cost function. Further details on the count of the Hadamard tests can be found in Appendix \ref{circuit_execution_appendix}.

One notable observation is that the number of gates for each typology, as well as the total count, appears to scale linearly with the number of qubits.
Additionally, there is a significant discrepancy between the number of gates for the Hadamard test circuit for the denominator and the numerator. The latter typically requires more gates, approximately five times more in the case of the \textit{ising} problem, and the difference is even more pronounced in the \textit{random pauli} problem, where it becomes exponentially larger.

Furthermore, for the computation of the denominator, there is a notable dependence on the specific problem being addressed. Whereas in the \textit{ising} problem with $n=10$ the number of gates is around 300, in the \textit{random pauli} problem with $n=10$ the total number of gates exceeds $10^7$. This significant difference between the two problems arises from the fact that in the \textit{random pauli} problem, the vector $\ket{b}$ is a general random real-valued vector with unitary norm, and therefore it is non-trivial to prepare in comparison with $\ket{b}$ in the \textit{ising} problem, which is an equal superposition of states and can be prepared with a single layer of Hadamard gates (and therefore $R_Y$ gates in terms of the basis used).

It is worth mentioning that for obtaining these results, we utilized Qiskit functions for both the state preparation of $\ket{b}$ in the \textit{random pauli} problem and for transpilation. While our approach provides valuable insights, it is important to acknowledge that there might be alternative methods or functions available to generate equivalent circuits with a lower number of gates.

\paragraph{Circuit depth}

\begin{figure}[htp]
    \centering
    \includegraphics[width=\columnwidth]{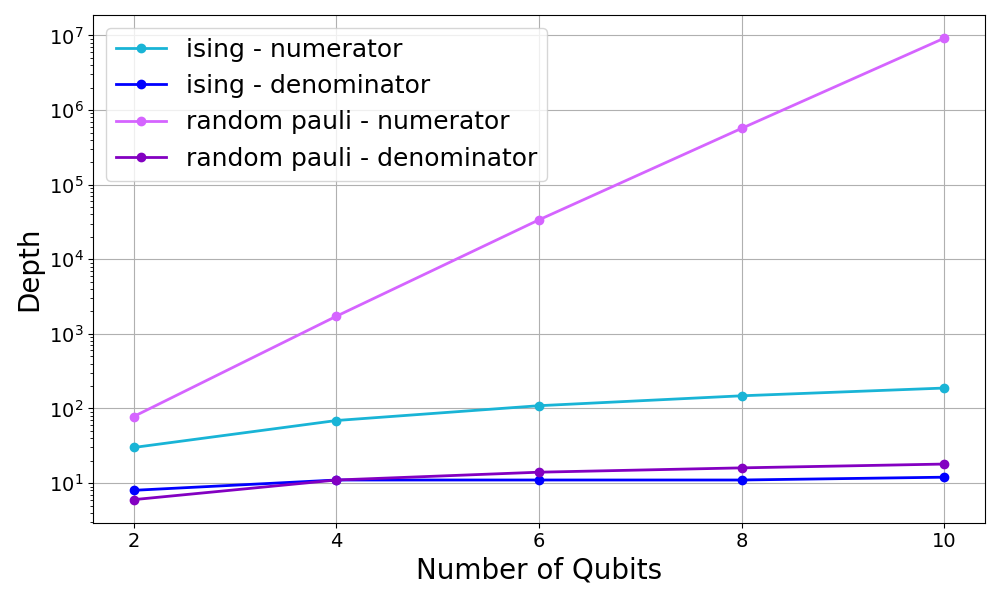}
    \caption{Average depths for Hadamard test circuits.}
    \label{fig:circuit_depth}
\end{figure}

Fig. \ref{fig:circuit_depth} illustrates the depths of distinct circuits for both \textit{ising} and \textit{random pauli} problem instances with $n=2, 4, 6, 8$, and $10$ qubits, presented in a logarithmic scale. Once again, we distinguish between the average circuit for the numerator and the average circuit for the denominator.

The figure highlights that the depth of the circuit required for the numerator is higher than the depth required for the denominator.
Moreover, there is a significant disparity in the required circuit depth for the numerator between \textit{ising} and \textit{random pauli} problems, with the latter demanding notably deeper circuits. This difference can be attributed to the contribution to circuit depth originated from the state preparation in the \textit{random pauli} problem. In the \textit{ising} problem, the single layer of Hadamard gates increases the circuit depth by only 1. In contrast, the preparation of the $\ket{b}$ state in the \textit{random pauli} problem requires a substantially larger number of gates, resulting in a circuit with increased depth.
All these observations on circuit depth are consistent with the findings on the gate count reported in the previous paragraph.

\subsection{Investigating the sub-optimal results} \label{sec:sub-optimal_results}

In this subsection, we explore the potential causes behind the inability to obtain satisfactory solutions when applying the VQLS algorithm to the \textit{random pauli} problem, contrasting the outcomes achieved with the \textit{ising} problem.
Our investigation centers on the cost function gradient, aiming to determine if the sub-optimal results may be due to a flat landscape of the cost function, which poses significant challenges for classical optimization.
Specifically, we begin our analysis by examining the presence of barren plateaus, and then we estimate the gradient at randomly selected points in the parameter space and at the points encountered during the optimization process.

\paragraph{Barren plateaus analysis}
Barren plateaus, as defined in \cite{arrasmith_2022}, is a phenomenon where the variance of all components of the gradient of the cost function decrease exponentially when the system size increases.
Therefore, to establish the existence of barren plateaus, we need to show the exponential decrease of the partial derivatives $\frac{\partial C(\theta)}{\partial \theta_i}$ for all $i=0, ..., m-1$, where $m$ denotes the number of parameters. This requires estimating the variance using:
\begin{equation}
\label{eq:variance}
    \text{Var}_{\theta}\left[\frac{\partial C(\theta)}{\partial \theta_i}\right]
    = \text{E}_{\theta}\left[\left(\frac{\partial C(\theta)}{\partial \theta_i}\right)^2\right]
    - \text{E}_{\theta}\left[\frac{\partial C(\theta)}{\partial \theta_i}\right]^2.
\end{equation}

In order to estimate $\text{E}_{\theta}\left[\frac{\partial C(\theta)}{\partial \theta_i}\right]$, we assume that each component $\theta_k$ of the parameter vector $\theta$ is independent and uniformly distributed in $[0, 4\pi]$. Therefore, the density of the distribution of $\theta$ is $\left(\frac{1}{4\pi}\right)^m \mathds{1}_{[0, 4\pi]^m}$, where $\mathds{1}_{[0, 4\pi]^m}$ denotes the indicator function over the multidimensional rectangular space $[0, 4\pi]^m$.
Additionally, due to the periodicity of the ansatz, both global and local cost functions satisfy $C(\theta_0,..., \theta_{i-1}, 0, \theta_{i+1},..., \theta_{m-1}) = C(\theta_0,..., \theta_{i-1}, 4\pi, \theta_{i+1},..., \theta_{m-1})$ for all $i=0,..., m-1$ and $\theta^{\bar{i}} \in [0,4\pi]^{m-1}$, where $\theta^{\bar{i}} \equiv (\theta_0,...,\theta_{i-1},\theta_{i+1},..., \theta_{m-1})$.

Then, we compute:
\begin{equation}
\begin{split}
    \text{E}_{\theta}\left[\frac{\partial C(\theta)}{\partial \theta_i}\right] 
    &=\left(\frac{1}{4 \pi}\right)^m\int_{[0,4\pi]^m} d\theta \frac{\partial C(\theta)}{\partial \theta_i} \\ 
    &=\left(\frac{1}{4 \pi}\right)^m\int_{[0,4\pi]^{m-1}} d\theta^{\bar{i}} \int_{[0,4\pi]} d\theta_i \frac{\partial C(\theta)}{\partial \theta_i} \\ 
    &= \left(\frac{1}{4 \pi}\right)^m\int_{[0,4\pi]^{m-1}} d\theta^{\bar{i}} \left[C(\theta)\Big|_{\theta_i=4\pi} - C(\theta)\Big|_{\theta_i=0}\right] \\
    &= 0.
\end{split}
\end{equation}

Note that the integral we are evaluating is a multidimensional integral over all parameters.
Therefore, we represent the differential volume elements as $d\theta \equiv \prod_{k=0}^{m-1} d\theta_k$ and $d\theta^{\bar{i}} \equiv \prod_{k=0}^{i-1} d\theta_k \prod_{k=i+1}^{m-1} d\theta_k$.
The third equality is derived from the fundamental theorem of calculus. Its application relies on the continuity and differentiability of the cost function $C$ for both the definitions in (\ref{eq:global_cost_function}) and (\ref{eq:local_cost_function}).
Indeed, when the matrix $A$ is non-singular (indicating a determined linear system, which is our case), the denominator in the calculation can never be equal to zero ensuring the continuity of the cost function. Moreover, the effect of the modulus in the numerator in (\ref{eq:global_cost_function}), which can potentially arise non-differentiability points, is nullified by the square term, ensuring differentiability.
Finally, the last equality follows from the periodicity of the cost function with respect to $\theta_i$.

Thus, according to \eqref{eq:variance}, to estimate the variance we need to evaluate:
\begin{equation}
\label{eq:expectation_square}
    \text{E}_{\theta}\left[\left(\frac{\partial C(\theta)}{\partial \theta_i}\right)^2\right] 
    = \left(\frac{1}{4\pi}\right)^m\int_{[0, 4\pi]^m} d \mathbf{\theta} \left(\frac{\partial C(\mathbf{\theta})}{\partial \theta_i}\right)^2
\end{equation}

However, for both the cost functions defined in (\ref{eq:global_cost_function}) and (\ref{eq:local_cost_function}), the integral in (\ref{eq:expectation_square}) cannot be computed analytically, due to the presence of the denominator $\braket{\psi}{\psi}$.
Therefore, we approximate (\ref{eq:expectation_square}) using a Monte Carlo algorithm.
Specifically, for each component $i$, we estimate the partial derivative of the cost function with respect to this component in 4096\footnote{In order to show that 4096 samples are sufficient, we performed the same variance estimates with double the number of samples. The new variance estimate exhibited minimal change and remained within the margin of error.} values of the parameter vector $\theta$, and then compute the average of the squared derivatives.
Our analysis focuses on the \textit{random pauli} problem, which includes 5 instances of varying sizes.

\begin{figure}[htp]
    \centering
    \includegraphics[width=\columnwidth]{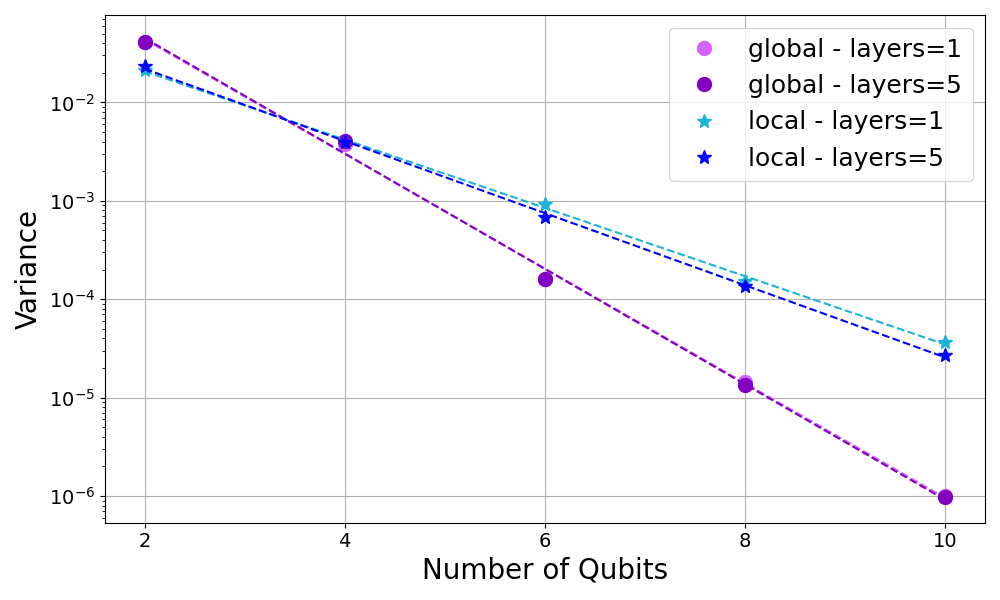}
    \caption{Variance of the first gradient component for the \textit{random Pauli} problem, estimated using a Monte Carlo method with 4096 samples, as a function of the number of qubits. Both global and local cost functions are analyzed, with ansatzes with 1 and 5 layers.}
    \label{fig:barren_plateaus}
\end{figure}

Fig. \ref{fig:barren_plateaus} shows the expected variance of the first gradient component as a function of the number of qubits for both global and local cost functions.
The choice of reporting the results only for the first component is motivated by the fact that the results obtained for the other components follow a similar trend.
Moreover, the trends of the variance obtained were consistent across layers, therefore we decided to report the results only for 1 and 5 layers. The results are presented on a logarithmic scale for the y axis, and the best fitting line for each case is also plotted.
Observing the figure, we can note an evident exponential decrease of the expected variance with the increasing number of qubits, indicating the presence of barren plateaus for this problem. Additionally, as expected from theoretical results, the exponential decrease is faster in the case of the global cost function than for the local one, which suggests that using the local cost function may reduce the effect of the barren plateau phenomenon.

\begin{table}[ht]
\centering
\caption{
Variance of the first gradient component for different numbers of layers for the \textit{random pauli} problem instance with $n=6$ qubits, using both global and local cost functions. \label{tab:gradient_variance_layers}} 
\resizebox{\linewidth}{!}{
\begin{tabular}{ccc}
    \toprule
        \textbf{layers} & \textbf{global} & \textbf{local}\\ 
        \midrule
        1 & $1.635 \cdot 10^{-4}$ ± $4.238 \cdot 10^{-7}$ & $9.290 \cdot 10^{-4}$ ± $1.269 \cdot 10^{-6}$ \\ 
        2 & $1.654 \cdot 10^{-4}$ ± $4.276 \cdot 10^{-7}$ & $8.261 \cdot 10^{-4}$ ± $1.174 \cdot 10^{-6}$ \\ 
        3 & $1.650 \cdot 10^{-4}$ ± $4.216 \cdot 10^{-7}$ & $7.815 \cdot 10^{-4}$ ± $1.123 \cdot 10^{-6}$ \\ 
        4 & $1.650 \cdot 10^{-4}$ ± $4.261 \cdot 10^{-7}$ & $7.453 \cdot 10^{-4}$ ± $1.081 \cdot 10^{-6}$ \\ 
        5 & $1.654 \cdot 10^{-4}$ ± $4.250 \cdot 10^{-7}$ & $7.189 \cdot 10^{-4}$ ± $1.043 \cdot 10^{-6}$ \\ 
    \bottomrule 
\end{tabular}
}
\end{table}

In Table \ref{tab:gradient_variance_layers} we report the variance of the first gradient component for different numbers of layers in the circuit for the \textit{random pauli} problem instance with $n=6$ qubits. The choice of considering a smaller number of qubits is motivated by the reduced computational cost necessary for estimating the variance of the gradient, which allows us to perform a higher number of samples, specifically 1,048,576, in a reasonable amount of time, resulting in higher precision in the estimation.
In the table, the variance is computed for both the global and the local cost functions. We observe that the trend for the global cost function is rather inconsistent, while the variance for the local cost function decreases consistently.

\paragraph{Estimation of the global cost function gradient}

\begin{figure*}[h!]
    \centering
    \begin{subfigure}{0.49\textwidth}
        \includegraphics[width=\linewidth]{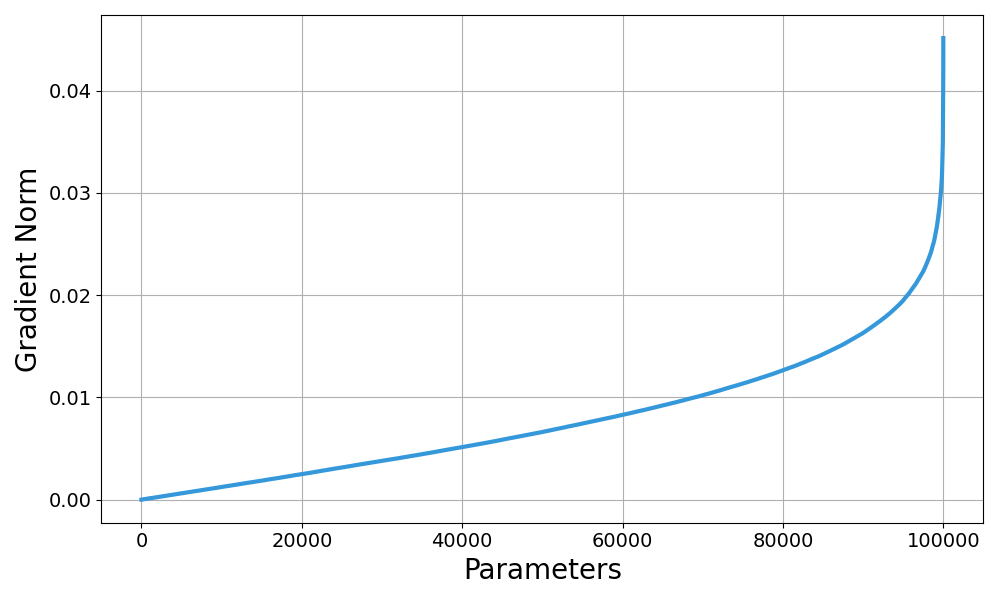}
        \caption{Gradient norm at random points. On the $x$ axis the indexes of the samples ordered from the ones having the smallest gradient norm to the ones having the highest gradient gradient norm.\label{fig:gradient_random_points}}
    \end{subfigure}\hfill
    \begin{subfigure}{0.49\textwidth}
        \includegraphics[width=\linewidth]{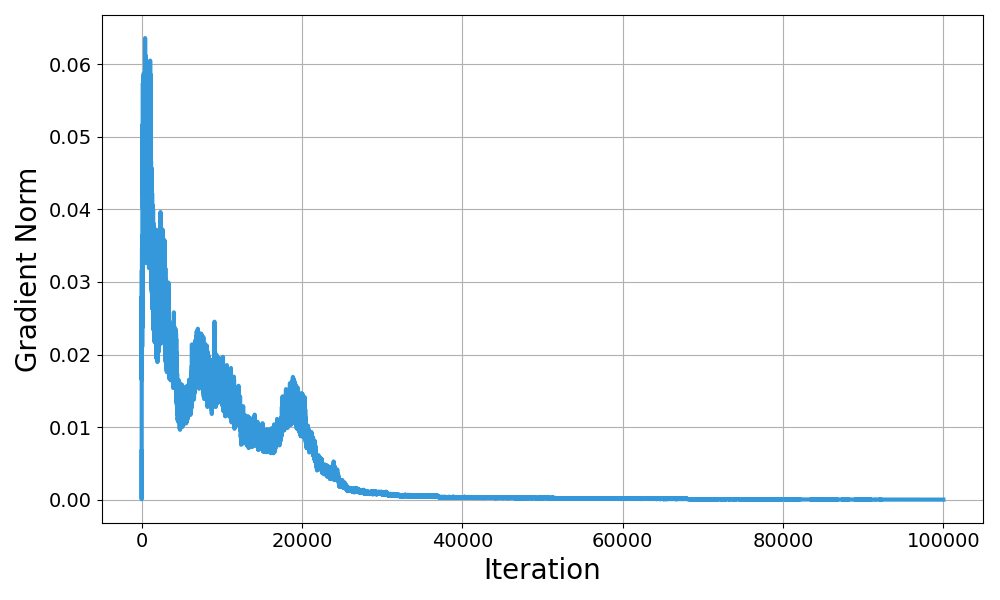}
        \caption{Gradient norm at the points explored during the algorithm execution. On the $x$ axis the indexes of the samples ordered based on the iterations in which they were explored.\label{fig:gradient_explored_points}}
    \end{subfigure}
    \caption{Gradient norm for the \textit{random pauli} problem with $n=10$ qubits and \textit{layers} = 5.\label{fig:gradient_norm}}
\end{figure*}

After analyzing the gradient variance, we aim to estimate the gradient of the cost function at specific points: some points randomly sampled in the parameter space and the points explored during the algorithm execution. We focus on the global cost function associated with a \textit{random pauli} problem instance involving $n=10$ qubits using an ansatz with 5 layers. This choice is motivated by previous studies \cite{holmes_2022} suggesting that the barren plateaus phenomenon is more pronounced with global cost functions and when the number of layers increases. Therefore, we consider a scenario where barren plateaus are most likely to occur.\footnote{The gradient is estimated using the Autograd library: \url{https://pytorch.org/docs/stable/generated/torch.autograd.grad.html}.}

Fig. \ref{fig:gradient_random_points} presents the norm of the gradient evaluated at 100,000 randomly sampled points, sorted from the smallest to the largest in norm. We observe that the gradient norm assumes non-zero values for most of the points, leading us to believe that the number of qubits considered is not high enough to encounter optimization issues connected to vanishing gradients due to the barren plateau phenomenon.
Additionally, Fig. \ref{fig:gradient_explored_points} shows the gradient norm at the points explored during the optimization process. We note that the gradient norm approaches zero during the optimization, suggesting that the optimizer can navigate effectively the cost function landscape.

However, despite the optimization appears to work correctly, the final results are not satisfactory, as the final cost function value is far from zero for the instances with $n=8$ and $n=10$ qubits. Since the results improve when the number of layers is increased, we believe that the sub-optimal results are likely due to the inadequate expressibility of the ansatz, which significantly impacts the algorithm capability of finding the correct solution. Nevertheless, as mentioned in Section \ref{sec:results_quality}, we do not further increase the number of layers, due to the prohibitively high computational cost.

\section{Conclusions}

In this study, we implemented the VQLS algorithm and conducted experiments across various problem instances, including instances with larger size and without specific structure, like those typically considered in previous literature.
We evaluated both the global and the local cost functions, simplifying their expressions to reduce the number of required circuit executions.

Our findings reveal that, while VQLS achieves high-quality results on \textit{ising} problem instances, it yields poor results on problems with more structured matrices, like \textit{random pauli} and \textit{darcy}. Moreover, for \textit{random pauli} instances, the results improve when increasing the number of layers in the ansatz, whereas for \textit{darcy} instances, the improvements are limited.
Additionally, we analyzed the convergence of the cost function and observed that it was able to reach 0 only for \textit{ising} instances. In all experiments, the cost function decreased, with the most pronounced decrease occurring during the initial iterations.
We also investigated the resources required for circuit implementation, noting the significant contribution of state preparation in increasing circuit depth and gate count.
We finally explored the reasons behind the sub-optimal results obtained with the \textit{random pauli} problem and showed that, even if the problem is prone to barren plateaus, this is probably not the cause of the sub-optimal results. 

These findings highlight the need for further research. While VQLS shows promising capabilities in efficiently encoding and solving large linear systems, achieving optimal results relies on finding an ansatz that contains the correct solution within the space of states that can be sampled and with trainable parameters.
Additionally, it is crucial to choose an appropriate ansatz and LCU decomposition for the linear system matrix, as well as an efficient sequence of gates for preparing the state \ket{b}, in order to make the circuit implementation feasible on real quantum hardware. 
Possible future research directions include analyzing the circuit expressibility with the goal of identifying suitable ansatzes for the problems at hand, elaborating more efficient LCU decompositions, and improving techniques for state preparation to reduce the amount of resources required for Hadamard tests.

Overall, we believe that the VQLS algorithm holds potential in solving specific large-scale linear systems, but only if its components, particularly the ansatz, are carefully chosen.

\section{Acknowledgment}
We acknowledge the financial support from ICSC - ``National Research Centre in High Performance Computing, Big Data and Quantum Computing'', funded by European Union – NextGenerationEU. 
We also acknowledge the support and computational resources provided by E4 Computer Engineering.
We acknowledge Giovanni Scrofani and Paolo Bozzoni from Eni SpA for providing the data to test the algorithm on the \textit{darcy} problem.
We also extend our appreciation to Claudio Sanavio from Istituto Italiano di Tecnologia for his insightful suggestions, which significantly improved the quality of this work.

\FloatBarrier

\bibliographystyle{IEEEtran}
\bibliography{IEEEabrv,bibliography}

\begin{thebibliography}{10}
\providecommand{\url}[1]{#1}
\csname url@samestyle\endcsname
\providecommand{\newblock}{\relax}
\providecommand{\bibinfo}[2]{#2}
\providecommand{\BIBentrySTDinterwordspacing}{\spaceskip=0pt\relax}
\providecommand{\BIBentryALTinterwordstretchfactor}{4}
\providecommand{\BIBentryALTinterwordspacing}{\spaceskip=\fontdimen2\font plus
\BIBentryALTinterwordstretchfactor\fontdimen3\font minus \fontdimen4\font\relax}
\providecommand{\BIBforeignlanguage}[2]{{%
\expandafter\ifx\csname l@#1\endcsname\relax
\typeout{** WARNING: IEEEtran.bst: No hyphenation pattern has been}%
\typeout{** loaded for the language `#1'. Using the pattern for}%
\typeout{** the default language instead.}%
\else
\language=\csname l@#1\endcsname
\fi
#2}}
\providecommand{\BIBdecl}{\relax}
\BIBdecl

\bibitem{preskill_2018}
\BIBentryALTinterwordspacing
J.~Preskill, ``Quantum {C}omputing in the {NISQ} era and beyond,'' \emph{{Quantum}}, vol.~2, p.~79, Aug. 2018. [Online]. Available: \url{https://doi.org/10.22331/q-2018-08-06-79}
\BIBentrySTDinterwordspacing

\bibitem{arute_2019}
\BIBentryALTinterwordspacing
F.~Arute, K.~Arya, R.~Babbush, D.~Bacon, J.~C. Bardin, R.~Barends, R.~Biswas, S.~Boixo, F.~G. S.~L. Brandao, D.~A. Buell, B.~Burkett, Y.~Chen, Z.~Chen, B.~Chiaro, R.~Collins, W.~Courtney, A.~Dunsworth, E.~Farhi, B.~Foxen, A.~Fowler, C.~Gidney, M.~Giustina, R.~Graff, K.~Guerin, S.~Habegger, M.~P. Harrigan, M.~J. Hartmann, A.~Ho, M.~Hoffmann, T.~Huang, T.~S. Humble, S.~V. Isakov, E.~Jeffrey, Z.~Jiang, D.~Kafri, K.~Kechedzhi, J.~Kelly, P.~V. Klimov, S.~Knysh, A.~Korotkov, F.~Kostritsa, D.~Landhuis, M.~Lindmark, E.~Lucero, D.~Lyakh, S.~Mandrà, J.~R. McClean, M.~McEwen, A.~Megrant, X.~Mi, K.~Michielsen, M.~Mohseni, J.~Mutus, O.~Naaman, M.~Neeley, C.~Neill, M.~Y. Niu, E.~Ostby, A.~Petukhov, J.~C. Platt, C.~Quintana, E.~G. Rieffel, P.~Roushan, N.~C. Rubin, D.~Sank, K.~J. Satzinger, V.~Smelyanskiy, K.~J. Sung, M.~D. Trevithick, A.~Vainsencher, B.~Villalonga, T.~White, Z.~J. Yao, P.~Yeh, A.~Zalcman, H.~Neven, and J.~M. Martinis, ``Quantum supremacy using a programmable superconducting processor,'' \emph{Nature}, vol.
  574, no. 7779, pp. 505--510, 2019. [Online]. Available: \url{https://doi.org/10.1038/s41586-019-1666-5}
\BIBentrySTDinterwordspacing

\bibitem{kim_2023}
\BIBentryALTinterwordspacing
Y.~Kim, A.~Eddins, S.~Anand, K.~X. Wei, E.~van~den Berg, S.~Rosenblatt, H.~Nayfeh, Y.~Wu, M.~Zaletel, K.~Temme, and A.~Kandala, ``Evidence for the utility of quantum computing before fault tolerance,'' \emph{Nature}, vol. 618, no. 7965, pp. 500--505, 2023. [Online]. Available: \url{https://doi.org/10.1038/s41586-023-06096-3}
\BIBentrySTDinterwordspacing

\bibitem{sood_2023}
S.~K. Sood and Pooja, ``Quantum computing review: A decade of research,'' \emph{IEEE Transactions on Engineering Management}, pp. 1--15, 2023.

\bibitem{daley_2022}
\BIBentryALTinterwordspacing
A.~J. Daley, I.~Bloch, C.~Kokail, S.~Flannigan, N.~Pearson, M.~Troyer, and P.~Zoller, ``Practical quantum advantage in quantum simulation,'' \emph{Nature}, vol. 607, no. 7920, pp. 667--676, 2022. [Online]. Available: \url{https://doi.org/10.1038/s41586-022-04940-6}
\BIBentrySTDinterwordspacing

\bibitem{franca_2021}
\BIBentryALTinterwordspacing
D.~Stilck~França and R.~García-Patrón, ``Limitations of optimization algorithms on noisy quantum devices,'' \emph{Nature Physics}, vol.~17, no.~11, pp. 1221--1227, 2021. [Online]. Available: \url{https://doi.org/10.1038/s41567-021-01356-3}
\BIBentrySTDinterwordspacing

\bibitem{pan_2023}
Z.~Pan, Y.~Feng, Z.~Li, Y.~Liu, and Y.~Li, ``Understanding the impact of quantum noise on quantum programs,'' in \emph{2023 IEEE International Conference on Software Analysis, Evolution and Reengineering (SANER)}, 2023, pp. 426--437.

\bibitem{fellous-asiani_2021}
\BIBentryALTinterwordspacing
M.~Fellous-Asiani, J.~H. Chai, R.~S. Whitney, A.~Auff\`eves, and H.~K. Ng, ``Limitations in quantum computing from resource constraints,'' \emph{PRX Quantum}, vol.~2, p. 040335, Nov 2021. [Online]. Available: \url{https://link.aps.org/doi/10.1103/PRXQuantum.2.040335}
\BIBentrySTDinterwordspacing

\bibitem{clerk_2010}
\BIBentryALTinterwordspacing
A.~A. Clerk, M.~H. Devoret, S.~M. Girvin, F.~Marquardt, and R.~J. Schoelkopf, ``Introduction to quantum noise, measurement, and amplification,'' \emph{Rev. Mod. Phys.}, vol.~82, pp. 1155--1208, Apr 2010. [Online]. Available: \url{https://link.aps.org/doi/10.1103/RevModPhys.82.1155}
\BIBentrySTDinterwordspacing

\bibitem{cerezo_2021_vqas}
\BIBentryALTinterwordspacing
M.~Cerezo, A.~Arrasmith, R.~Babbush, S.~C. Benjamin, S.~Endo, K.~Fujii, J.~R. McClean, K.~Mitarai, X.~Yuan, L.~Cincio, and P.~J. Coles, ``Variational quantum algorithms,'' \emph{Nature Reviews Physics}, vol.~3, no.~9, pp. 625--644, 2021. [Online]. Available: \url{https://doi.org/10.1038/s42254-021-00348-9}
\BIBentrySTDinterwordspacing

\bibitem{qin_2023}
\BIBentryALTinterwordspacing
J.~Qin, ``Review of ansatz designing techniques for variational quantum algorithms,'' \emph{Journal of Physics: Conference Series}, vol. 2634, no.~1, p. 012043, nov 2023. [Online]. Available: \url{https://dx.doi.org/10.1088/1742-6596/2634/1/012043}
\BIBentrySTDinterwordspacing

\bibitem{wurtz_2021}
J.~Wurtz and P.~J. Love, ``Classically optimal variational quantum algorithms,'' \emph{IEEE Transactions on Quantum Engineering}, vol.~2, pp. 1--7, 2021.

\bibitem{sim_2019}
\BIBentryALTinterwordspacing
S.~Sim, P.~D. Johnson, and A.~Aspuru-Guzik, ``Expressibility and entangling capability of parameterized quantum circuits for hybrid quantum-classical algorithms,'' \emph{Advanced Quantum Technologies}, vol.~2, no.~12, p. 1900070, 2019. [Online]. Available: \url{https://onlinelibrary.wiley.com/doi/abs/10.1002/qute.201900070}
\BIBentrySTDinterwordspacing

\bibitem{du_2020}
\BIBentryALTinterwordspacing
Y.~Du, M.-H. Hsieh, T.~Liu, and D.~Tao, ``Expressive power of parametrized quantum circuits,'' \emph{Phys. Rev. Res.}, vol.~2, p. 033125, Jul 2020. [Online]. Available: \url{https://link.aps.org/doi/10.1103/PhysRevResearch.2.033125}
\BIBentrySTDinterwordspacing

\bibitem{mcclean_2018}
\BIBentryALTinterwordspacing
J.~R. McClean, S.~Boixo, V.~N. Smelyanskiy, R.~Babbush, and H.~Neven, ``Barren plateaus in quantum neural network training landscapes,'' \emph{Nature Communications}, vol.~9, no.~1, Nov. 2018. [Online]. Available: \url{http://dx.doi.org/10.1038/s41467-018-07090-4}
\BIBentrySTDinterwordspacing

\bibitem{arrasmith_2021}
\BIBentryALTinterwordspacing
A.~Arrasmith, M.~Cerezo, P.~Czarnik, L.~Cincio, and P.~J. Coles, ``Effect of barren plateaus on gradient-free optimization,'' \emph{{Quantum}}, vol.~5, p. 558, Oct. 2021. [Online]. Available: \url{https://doi.org/10.22331/q-2021-10-05-558}
\BIBentrySTDinterwordspacing

\bibitem{arrasmith_2022}
\BIBentryALTinterwordspacing
A.~Arrasmith, Z.~Holmes, M.~Cerezo, and P.~J. Coles, ``Equivalence of quantum barren plateaus to cost concentration and narrow gorges,'' \emph{Quantum Science and Technology}, vol.~7, no.~4, p. 045015, aug 2022. [Online]. Available: \url{https://dx.doi.org/10.1088/2058-9565/ac7d06}
\BIBentrySTDinterwordspacing

\bibitem{cerezo_2021}
\BIBentryALTinterwordspacing
M.~Cerezo and P.~J. Coles, ``Higher order derivatives of quantum neural networks with barren plateaus,'' \emph{Quantum Science and Technology}, vol.~6, no.~3, p. 035006, jun 2021. [Online]. Available: \url{https://dx.doi.org/10.1088/2058-9565/abf51a}
\BIBentrySTDinterwordspacing

\bibitem{holmes_2022}
\BIBentryALTinterwordspacing
Z.~Holmes, K.~Sharma, M.~Cerezo, and P.~J. Coles, ``Connecting ansatz expressibility to gradient magnitudes and barren plateaus,'' \emph{PRX Quantum}, vol.~3, p. 010313, Jan 2022. [Online]. Available: \url{https://link.aps.org/doi/10.1103/PRXQuantum.3.010313}
\BIBentrySTDinterwordspacing

\bibitem{larocca_2022}
\BIBentryALTinterwordspacing
M.~Larocca, P.~Czarnik, K.~Sharma, G.~Muraleedharan, P.~J. Coles, and M.~Cerezo, ``Diagnosing {B}arren {P}lateaus with {T}ools from {Q}uantum {O}ptimal {C}ontrol,'' \emph{{Quantum}}, vol.~6, p. 824, Sep. 2022. [Online]. Available: \url{https://doi.org/10.22331/q-2022-09-29-824}
\BIBentrySTDinterwordspacing

\bibitem{volkoff_2021}
\BIBentryALTinterwordspacing
T.~Volkoff and P.~J. Coles, ``Large gradients via correlation in random parameterized quantum circuits,'' \emph{Quantum Science and Technology}, vol.~6, no.~2, p. 025008, jan 2021. [Online]. Available: \url{https://dx.doi.org/10.1088/2058-9565/abd891}
\BIBentrySTDinterwordspacing

\bibitem{sanavio_2024}
\BIBentryALTinterwordspacing
C.~Sanavio, S.~Tibaldi, E.~Tignone, and E.~Ercolessi, ``Quantum circuit for imputation of missing data,'' 2024. [Online]. Available: \url{https://arxiv.org/abs/2405.04367}
\BIBentrySTDinterwordspacing

\bibitem{peruzzo_2014}
\BIBentryALTinterwordspacing
A.~Peruzzo, J.~McClean, P.~Shadbolt, M.-H. Yung, X.-Q. Zhou, P.~J. Love, A.~Aspuru-Guzik, and J.~L. O’Brien, ``A variational eigenvalue solver on a photonic quantum processor,'' \emph{Nature Communications}, vol.~5, no.~1, Jul. 2014. [Online]. Available: \url{http://dx.doi.org/10.1038/ncomms5213}
\BIBentrySTDinterwordspacing

\bibitem{tilly_2021}
\BIBentryALTinterwordspacing
J.~Tilly, H.~Chen, S.~Cao, D.~Picozzi, K.~Setia, Y.~Li, E.~Grant, L.~Wossnig, I.~Rungger, G.~H. Booth, and J.~Tennyson, ``The variational quantum eigensolver: A review of methods and best practices,'' \emph{Physics Reports}, vol. 986, pp. 1--128, 2022, the Variational Quantum Eigensolver: a review of methods and best practices. [Online]. Available: \url{https://www.sciencedirect.com/science/article/pii/S0370157322003118}
\BIBentrySTDinterwordspacing

\bibitem{cao_2019}
\BIBentryALTinterwordspacing
Y.~Cao, J.~Romero, J.~P. Olson, M.~Degroote, P.~D. Johnson, M.~Kieferová, I.~D. Kivlichan, T.~Menke, B.~Peropadre, N.~P.~D. Sawaya, S.~Sim, L.~Veis, and A.~Aspuru-Guzik, ``Quantum chemistry in the age of quantum computing,'' \emph{Chemical Reviews}, vol. 119, no.~19, pp. 10\,856--10\,915, 2019, pMID: 31469277. [Online]. Available: \url{https://doi.org/10.1021/acs.chemrev.8b00803}
\BIBentrySTDinterwordspacing

\bibitem{mcclean_2016}
\BIBentryALTinterwordspacing
J.~R. McClean, J.~Romero, R.~Babbush, and A.~Aspuru-Guzik, ``The theory of variational hybrid quantum-classical algorithms,'' \emph{New Journal of Physics}, vol.~18, no.~2, p. 023023, feb 2016. [Online]. Available: \url{https://dx.doi.org/10.1088/1367-2630/18/2/023023}
\BIBentrySTDinterwordspacing

\bibitem{farhi_2014}
E.~Farhi, J.~Goldstone, and S.~Gutmann, ``A quantum approximate optimization algorithm,'' 2014.

\bibitem{blekos_2023}
K.~Blekos, D.~Brand, A.~Ceschini, C.-H. Chou, R.-H. Li, K.~Pandya, and A.~Summer, ``A review on quantum approximate optimization algorithm and its variants,'' 2023.

\bibitem{crooks_2018}
G.~E. Crooks, ``Performance of the quantum approximate optimization algorithm on the maximum cut problem,'' 2018.

\bibitem{wang_2018}
\BIBentryALTinterwordspacing
Z.~Wang, S.~Hadfield, Z.~Jiang, and E.~G. Rieffel, ``Quantum approximate optimization algorithm for {MaxCut}: A fermionic view,'' \emph{Physical Review A}, vol.~97, no.~2, feb 2018. [Online]. Available: \url{https://doi.org/10.1103\%2Fphysreva.97.022304}
\BIBentrySTDinterwordspacing

\bibitem{basso_2022}
\BIBentryALTinterwordspacing
J.~Basso, E.~Farhi, K.~Marwaha, B.~Villalonga, and L.~Zhou, ``\BIBforeignlanguage{en}{The quantum approximate optimization algorithm at high depth for maxcut on large-girth regular graphs and the sherrington-kirkpatrick model},'' in \emph{\BIBforeignlanguage{en}{Theory of Quantum Computation, Communication, and Cryptography}}.\hskip 1em plus 0.5em minus 0.4em\relax Schloss Dagstuhl - Leibniz-Zentrum für Informatik, 2022. [Online]. Available: \url{https://drops.dagstuhl.de/opus/volltexte/2022/16514/}
\BIBentrySTDinterwordspacing

\bibitem{farhi_2017}
E.~Farhi, J.~Goldstone, S.~Gutmann, and H.~Neven, ``Quantum algorithms for fixed qubit architectures,'' 2017.

\bibitem{shaydulin_2023}
\BIBentryALTinterwordspacing
R.~Shaydulin, P.~C. Lotshaw, J.~Larson, J.~Ostrowski, and T.~S. Humble, ``Parameter transfer for quantum approximate optimization of weighted {MaxCut},'' \emph{{ACM} Transactions on Quantum Computing}, vol.~4, no.~3, pp. 1--15, apr 2023. [Online]. Available: \url{https://doi.org/10.1145\%2F3584706}
\BIBentrySTDinterwordspacing

\bibitem{stechly_2023}
M.~Stęchły, L.~Gao, B.~Yogendran, E.~Fontana, and M.~Rudolph, ``Connecting the hamiltonian structure to the qaoa energy and fourier landscape structure,'' 2023.

\bibitem{zhou_2020}
\BIBentryALTinterwordspacing
L.~Zhou, S.-T. Wang, S.~Choi, H.~Pichler, and M.~D. Lukin, ``Quantum approximate optimization algorithm: Performance, mechanism, and implementation on near-term devices,'' \emph{Physical Review X}, vol.~10, no.~2, jun 2020. [Online]. Available: \url{https://doi.org/10.1103\%2Fphysrevx.10.021067}
\BIBentrySTDinterwordspacing

\bibitem{larkin_2022}
\BIBentryALTinterwordspacing
J.~Larkin, M.~Jonsson, D.~Justice, and G.~G. Guerreschi, ``Evaluation of {QAOA} based on the approximation ratio of individual samples,'' \emph{Quantum Science and Technology}, vol.~7, no.~4, p. 045014, aug 2022. [Online]. Available: \url{https://doi.org/10.1088\%2F2058-9565\%2Fac6973}
\BIBentrySTDinterwordspacing

\bibitem{cook_2020}
J.~Cook, S.~Eidenbenz, and A.~Bärtschi, ``The quantum alternating operator ansatz on maximum k-vertex cover,'' in \emph{2020 IEEE International Conference on Quantum Computing and Engineering (QCE)}, 2020, pp. 83--92.

\bibitem{tabi_2020_graph_colouring}
\BIBentryALTinterwordspacing
Z.~Tabi, K.~H. El-Safty, Z.~Kallus, P.~Haga, T.~Kozsik, A.~Glos, and Z.~Zimboras, ``Quantum optimization for the graph coloring problem with space-efficient embedding,'' in \emph{2020 {IEEE} International Conference on Quantum Computing and Engineering ({QCE})}.\hskip 1em plus 0.5em minus 0.4em\relax {IEEE}, oct 2020. [Online]. Available: \url{https://doi.org/10.1109\%2Fqce49297.2020.00018}
\BIBentrySTDinterwordspacing

\bibitem{lin_2016_csp}
C.~Y.-Y. Lin and Y.~Zhu, ``Performance of qaoa on typical instances of constraint satisfaction problems with bounded degree,'' 2016.

\bibitem{willsch_2020}
\BIBentryALTinterwordspacing
M.~Willsch, D.~Willsch, F.~Jin, H.~D. Raedt, and K.~Michielsen, ``Benchmarking the quantum approximate optimization algorithm,'' \emph{Quantum Information Processing}, vol.~19, no.~7, jun 2020. [Online]. Available: \url{https://doi.org/10.1007\%2Fs11128-020-02692-8}
\BIBentrySTDinterwordspacing

\bibitem{yan_2022_factoring}
B.~Yan, Z.~Tan, S.~Wei, H.~Jiang, W.~Wang, H.~Wang, L.~Luo, Q.~Duan, Y.~Liu, W.~Shi, Y.~Fei, X.~Meng, Y.~Han, Z.~Shan, J.~Chen, X.~Zhu, C.~Zhang, F.~Jin, H.~Li, C.~Song, Z.~Wang, Z.~Ma, H.~Wang, and G.-L. Long, ``Factoring integers with sublinear resources on a superconducting quantum processor,'' 2022.

\bibitem{radzihovsky_2019_tsp}
M.~Radzihovsky, J.~Murphy, and M.~Swofford, ``A qaoa solution to the traveling salesman problem using pyquil,'' 2019.

\bibitem{brandhofer_2022_portfolio_opt}
\BIBentryALTinterwordspacing
S.~Brandhofer, D.~Braun, V.~Dehn, G.~Hellstern, M.~H{\"u}ls, Y.~Ji, I.~Polian, A.~S. Bhatia, and T.~Wellens, ``Benchmarking the performance of portfolio optimization with qaoa,'' \emph{Quantum Information Processing}, vol.~22, no.~1, p.~25, Dec 2022. [Online]. Available: \url{https://doi.org/10.1007/s11128-022-03766-5}
\BIBentrySTDinterwordspacing

\bibitem{kurowski_2023_jssp}
\BIBentryALTinterwordspacing
K.~Kurowski, T.~Pecyna, M.~Slysz, R.~Różycki, G.~Waligóra, and J.~W\k{e}glarz, ``Application of quantum approximate optimization algorithm to job shop scheduling problem,'' \emph{European Journal of Operational Research}, vol. 310, no.~2, pp. 518--528, 2023. [Online]. Available: \url{https://www.sciencedirect.com/science/article/pii/S0377221723002072}
\BIBentrySTDinterwordspacing

\bibitem{bravo-prieto_2023}
\BIBentryALTinterwordspacing
C.~Bravo{-}Prieto, R.~Larose, M.~Cerezo, Y.~Subasi, L.~Cincio, and P.~J. Coles, ``Variational quantum linear solver,'' \emph{Quantum}, vol.~7, p. 1188, 2023. [Online]. Available: \url{https://doi.org/10.22331/q-2023-11-22-1188}
\BIBentrySTDinterwordspacing

\bibitem{harrow_2009_hhl}
\BIBentryALTinterwordspacing
A.~W. Harrow, A.~Hassidim, and S.~Lloyd, ``Quantum algorithm for linear systems of equations,'' \emph{Phys. Rev. Lett.}, vol. 103, p. 150502, Oct 2009. [Online]. Available: \url{https://link.aps.org/doi/10.1103/PhysRevLett.103.150502}
\BIBentrySTDinterwordspacing

\bibitem{zaman_2023_hhl}
\BIBentryALTinterwordspacing
A.~Zaman, H.~J. Morrell, and W.~H. Yung, ``A step-by-step {HHL} algorithm walkthrough to enhance understanding of critical quantum computing concepts,'' \emph{{IEEE} Access}, vol.~11, pp. 77\,117--77\,131, 2023. [Online]. Available: \url{https://doi.org/10.1109/ACCESS.2023.3297658}
\BIBentrySTDinterwordspacing

\bibitem{schuld_2018}
\BIBentryALTinterwordspacing
M.~Schuld and F.~Petruccione, \emph{Information Encoding}.\hskip 1em plus 0.5em minus 0.4em\relax Cham: Springer International Publishing, 2018, pp. 139--171. [Online]. Available: \url{https://doi.org/10.1007/978-3-319-96424-9_5}
\BIBentrySTDinterwordspacing

\bibitem{liu_2024}
\BIBentryALTinterwordspacing
Y.~Liu, Z.~Chen, C.~Shu, P.~Rebentrost, Y.~Liu, S.~Chew, B.~Khoo, and Y.~Cui, ``A variational quantum algorithm-based numerical method for solving potential and stokes flows,'' \emph{Ocean Engineering}, vol. 292, p. 116494, 2024. [Online]. Available: \url{https://www.sciencedirect.com/science/article/pii/S0029801823028780}
\BIBentrySTDinterwordspacing

\bibitem{trahan_2023}
\BIBentryALTinterwordspacing
C.~J. Trahan, M.~Loveland, N.~Davis, and E.~Ellison, ``A variational quantum linear solver application to discrete finite-element methods,'' \emph{Entropy}, vol.~25, no.~4, p. 580, 2023. [Online]. Available: \url{https://doi.org/10.3390/e25040580}
\BIBentrySTDinterwordspacing

\bibitem{liu_2022}
\BIBentryALTinterwordspacing
Y.~Y. Liu, Z.~Chen, C.~Shu, S.~C. Chew, B.~C. Khoo, X.~Zhao, and Y.~D. Cui, ``{Application of a variational hybrid quantum-classical algorithm to heat conduction equation and analysis of time complexity},'' \emph{Physics of Fluids}, vol.~34, no.~11, p. 117121, 11 2022. [Online]. Available: \url{https://doi.org/10.1063/5.0121778}
\BIBentrySTDinterwordspacing

\bibitem{shang_2023}
\BIBentryALTinterwordspacing
R.~Shang, Z.~Wang, S.~Shi, J.~Li, Y.~Li, and Y.~Gu, ``Algorithm for simulating ocean circulation on a quantum computer,'' \emph{Science China Earth Sciences}, vol.~66, no.~10, pp. 2254--2264, October 2023. [Online]. Available: \url{https://doi.org/10.1007/s11430-023-1162-x}
\BIBentrySTDinterwordspacing

\bibitem{luo_2024}
\BIBentryALTinterwordspacing
H.~Luo, Q.~Zhou, Z.~Li, and Y.~Deng, ``Variational quantum linear solver-based combination rules in dempster–shafer theory,'' \emph{Information Fusion}, vol. 102, p. 102070, 2024. [Online]. Available: \url{https://www.sciencedirect.com/science/article/pii/S156625352300386X}
\BIBentrySTDinterwordspacing

\bibitem{xing_2023}
\BIBentryALTinterwordspacing
X.~Xing, A.~Gomez~Cadavid, A.~F. Izmaylov, and T.~V. Tscherbul, ``A hybrid quantum-classical algorithm for multichannel quantum scattering of atoms and molecules,'' \emph{The Journal of Physical Chemistry Letters}, vol.~14, no.~27, pp. 6224--6233, 2023. [Online]. Available: \url{https://doi.org/10.1021/acs.jpclett.3c00985}
\BIBentrySTDinterwordspacing

\bibitem{ali_2023}
M.~Ali and M.~Kabel, ``Performance study of variational quantum algorithms for solving the poisson equation on a quantum computer,'' \emph{Physical Review Applied}, vol.~20, no.~1, p. 014054, 2023.

\bibitem{patil_2022}
\BIBentryALTinterwordspacing
H.~Patil, Y.~Wang, and P.~S. Krsti\ifmmode~\acute{c}\else \'{c}\fi{}, ``Variational quantum linear solver with a dynamic ansatz,'' \emph{Phys. Rev. A}, vol. 105, p. 012423, Jan 2022. [Online]. Available: \url{https://link.aps.org/doi/10.1103/PhysRevA.105.012423}
\BIBentrySTDinterwordspacing

\bibitem{huang_2021}
\BIBentryALTinterwordspacing
H.-Y. Huang, K.~Bharti, and P.~Rebentrost, ``Near-term quantum algorithms for linear systems of equations with regression loss functions,'' \emph{New Journal of Physics}, vol.~23, no.~11, p. 113021, nov 2021. [Online]. Available: \url{https://dx.doi.org/10.1088/1367-2630/ac325f}
\BIBentrySTDinterwordspacing

\bibitem{yi_2023}
\BIBentryALTinterwordspacing
J.~Yi, K.~Suresh, A.~Moghiseh, and N.~Wehn, ``Variational quantum linear solver enhanced quantum support vector machine,'' \emph{CoRR}, vol. abs/2309.07770, 2023. [Online]. Available: \url{https://doi.org/10.48550/arXiv.2309.07770}
\BIBentrySTDinterwordspacing

\bibitem{pellow-jarman_2023}
\BIBentryALTinterwordspacing
A.~Pellow-Jarman, I.~Sinayskiy, A.~Pillay, and F.~Petruccione, ``Near term algorithms for linear systems of equations,'' \emph{Quantum Information Processing}, vol.~22, no.~6, p. 258, June 24 2023. [Online]. Available: \url{https://doi.org/10.1007/s11128-023-04020-2}
\BIBentrySTDinterwordspacing

\bibitem{saito_2023}
Y.~Saito, X.~Lee, D.~Cai, J.~Shin, and N.~Asai, ``Iterative refinement for variational quantum linear solver,'' in \emph{Proceedings of International Conference on Data Analytics and Insights, ICDAI 2023}, N.~Chaki, N.~D. Roy, P.~Debnath, and K.~Saeed, Eds.\hskip 1em plus 0.5em minus 0.4em\relax Singapore: Springer Nature Singapore, 2023, pp. 15--27.

\bibitem{pellow-jarman_2021}
\BIBentryALTinterwordspacing
A.~Pellow-Jarman, I.~Sinayskiy, A.~Pillay, and F.~Petruccione, ``A comparison of various classical optimizers for a variational quantum linear solver,'' \emph{Quantum Information Processing}, vol.~20, no.~6, jun 2021. [Online]. Available: \url{https://doi.org/10.1007/s11128-021-03140-x}
\BIBentrySTDinterwordspacing

\bibitem{liu_2021}
\BIBentryALTinterwordspacing
H.-L. Liu, Y.-S. Wu, L.-C. Wan, S.-J. Pan, S.-J. Qin, F.~Gao, and Q.-Y. Wen, ``Variational quantum algorithm for the poisson equation,'' \emph{Phys. Rev. A}, vol. 104, p. 022418, Aug 2021. [Online]. Available: \url{https://link.aps.org/doi/10.1103/PhysRevA.104.022418}
\BIBentrySTDinterwordspacing

\bibitem{xu_2021}
\BIBentryALTinterwordspacing
X.~Xu, J.~Sun, S.~Endo, Y.~Li, S.~C. Benjamin, and X.~Yuan, ``Variational algorithms for linear algebra,'' \emph{Science Bulletin}, vol.~66, no.~21, pp. 2181--2188, 2021. [Online]. Available: \url{https://www.sciencedirect.com/science/article/pii/S2095927321004631}
\BIBentrySTDinterwordspacing

\bibitem{cappanera_2021}
\BIBentryALTinterwordspacing
E.~Cappanera, ``Variational quantum linear solver for finite element problems: a poisson equation test case,'' master thesis, Delft University of Technology, 2021. [Online]. Available: \url{http://resolver.tudelft.nl/uuid:deba389d-f30f-406c-ad7b-babb1b298d87}
\BIBentrySTDinterwordspacing

\bibitem{cerezo_2021_bp_local}
\BIBentryALTinterwordspacing
M.~Cerezo, A.~Sone, T.~Volkoff, L.~Cincio, and P.~J. Coles, ``Cost function dependent barren plateaus in shallow parametrized quantum circuits,'' \emph{Nature Communications}, vol.~12, no.~1, p. 1791, 03 2021. [Online]. Available: \url{https://doi.org/10.1038/s41467-021-21728-w}
\BIBentrySTDinterwordspacing

\bibitem{sato_2021}
Y.~Sato, R.~Kondo, S.~Koide, H.~Takamatsu, and N.~Imoto, ``Variational quantum algorithm based on the minimum potential energy for solving the poisson equation,'' \emph{Physical Review A}, vol. 104, no.~5, p. 052409, 2021.

\bibitem{hadfield_2019}
\BIBentryALTinterwordspacing
S.~Hadfield, Z.~Wang, B.~O’Gorman, E.~Rieffel, D.~Venturelli, and R.~Biswas, ``From the quantum approximate optimization algorithm to a quantum alternating operator ansatz,'' \emph{Algorithms}, vol.~12, no.~2, p.~34, Feb. 2019. [Online]. Available: \url{http://dx.doi.org/10.3390/a12020034}
\BIBentrySTDinterwordspacing

\bibitem{hantschel_2009}
T.~Hantschel and A.~I. Kauerauf, \emph{Fundamentals of basin and petroleum systems modeling}.\hskip 1em plus 0.5em minus 0.4em\relax Springer Science \& Business Media, 2009.

\bibitem{powell_1994_cobyla}
\BIBentryALTinterwordspacing
M.~J.~D. Powell, \emph{A Direct Search Optimization Method That Models the Objective and Constraint Functions by Linear Interpolation}.\hskip 1em plus 0.5em minus 0.4em\relax Dordrecht: Springer Netherlands, 1994, pp. 51--67. [Online]. Available: \url{https://doi.org/10.1007/978-94-015-8330-5_4}
\BIBentrySTDinterwordspacing

\bibitem{singh_2023_cobyla2}
\BIBentryALTinterwordspacing
H.~Singh, S.~Majumder, and S.~Mishra, ``{Benchmarking of different optimizers in the variational quantum algorithms for applications in quantum chemistry},'' \emph{The Journal of Chemical Physics}, vol. 159, no.~4, p. 044117, 07 2023. [Online]. Available: \url{https://doi.org/10.1063/5.0161057}
\BIBentrySTDinterwordspacing

\bibitem{fernandez-pendas_2020_cobyla3}
\BIBentryALTinterwordspacing
M.~Fernández-Pendás, E.~F. Combarro, S.~Vallecorsa, J.~Ranilla, and I.~F. Rúa, ``A study of the performance of classical minimizers in the quantum approximate optimization algorithm,'' \emph{Journal of Computational and Applied Mathematics}, vol. 404, p. 113388, 2022. [Online]. Available: \url{https://www.sciencedirect.com/science/article/pii/S0377042721000078}
\BIBentrySTDinterwordspacing

\end{thebibliography}

\FloatBarrier

\appendix

\subsection{Hadamard test} \label{hadamard_test_appendix}

The Hadamard test is a quantum algorithm that estimates the real part of the expectation $\bra{\psi}U\ket{\psi}$, where $U$ is a unitary operator and \ket{\psi} is a quantum state. The procedure involves preparing the state \ket{\psi} on an $n$-qubit register and introducing a separate ancillary qubit. A Hadamard gate is applied to the ancilla, followed by a controlled-$U$ operation with the ancilla as the control qubit. Finally, another Hadamard gate is applied to the ancillary qubit (see Fig. \ref{fig:hadamard_test_circuit}). The probability of measuring the ancillary qubit in the state $\ket{1}$ provides an estimate of the real part of the expectation value, given by
\begin{equation}
\Re{\bra{\psi} U \ket{\psi}} = 1 - 2P(1).
\end{equation}
Here, $P(1)$ is the probability of measuring the ancillary qubit in the state \ket{1}.

In order to estimate the imaginary part of the expectation value, a modified version of the Hadamard test circuit can be employed. Specifically, after the first Hadamard gate, an $S^\dag$ gate is introduced before proceeding with the controlled-$U$ operation. This modification is illustrated in Fig. \ref{fig:hadamard_test_circuit}, where the new $S^\dag$ gate is in a red dashed box. The resulting probability of measuring the ancillary qubit in the state $\ket{1}$ can then be used to estimate the imaginary part of the expectation as follows:
\begin{equation}
\Im{\bra{\psi} U \ket{\psi}} = 1 - 2P(1).
\end{equation}

\begin{figure}
\centering
\caption{Hadamard test circuit for computing the real and imaginary parts of the expectation of the operator $U$ on the state \ket{\psi}. The $S^\dag$ gate (in the red dashed box) should be included only when estimating the imaginary part of the expectation.\label{fig:hadamard_test_circuit}}
    \begin{quantikz}[wire types={q,q}]
        \lstick{$\ket{0}$} & \gate{H} & \gate{S^\dag}\gategroup[1,steps=1,style={dashed, fill=red!20, inner xsep=2pt},background]{} & \ctrl{1} & \gate{H} & \meter{} \\
        \lstick{$\ket{\psi}_n$} & \qwbundle{n} & & \gate{U} & & 
    \end{quantikz}
\end{figure}

\subsection{Cost Function Computation} \label{cost_function_appendix}

Here we explain how to obtain the expressions (\ref{eq:real_glob_den}), (\ref{eq:real_glob_num}), and (\ref{eq:real_loc_num}) for computing the values of the global and the local cost functions.

\paragraph{Global Cost Function}

Computing the global cost function (\ref{eq:global_cost_function}) involves estimating the terms $\braket{\psi}{\psi}$ and $\abs{\braket{b}{\psi}}^2$.

Let us illustrate the derivation of $\braket{\psi}{\psi}$.
Given the LCU decomposition (\ref{eq:LCU_decomposition}) of the matrix $A$, we express $\braket{\psi}{\psi}$ as follows:
\begin{equation}
    \begin{split}
        \braket{\psi}{\psi} 
        &= \bra{0} V^\dag A^\dag A V \ket{0} 
        = \sum_{i=0}^{L-1} \sum_{j=0}^{L-1} c_i^* c_j \bra{0} V^\dag A_i^\dag A_j V \ket{0}.
    \end{split}
\end{equation}

To simplify the expression, we split the sum and observe that the terms can be paired, allowing us to consider a reduced number of terms:
\begin{equation}
    \label{eq:glob_den1}
    \begin{split}
        \braket{\psi}{\psi}
        &= \sum_{i=0}^{L-1}c_i^* c_i \bra{0} V^\dag A_i^\dag A_i V \ket{0} \\
        &\quad+ \sum_{i=0}^{L-1} \sum_{j=i+1}^{L-1} c_i^* c_j \bra{0} V^\dag A_i^\dag A_j V \ket{0} \\
        &\quad+ \sum_{i=0}^{L-1} \sum_{j=0}^{i-1} c_i^* c_j \bra{0} V^\dag A_i^\dag A_j V \ket{0}\\
        &= \sum_{i=0}^{L-1} \abs{c_i}^2 
        + \sum_{i=0}^{L-1} \sum_{j=i+1}^{L-1} c_i^* c_j \bra{0} V^\dag A_i^\dag A_j V \ket{0} \\
        &\quad+ \sum_{i=0}^{L-1} \sum_{j=i+1}^{L-1} \overline{c_i^* c_j \bra{0} V^\dag A_i^\dag A_j V \ket{0}} \\
        &= \sum_{i=0}^{L-1} \abs{c_i}^2 
        + \sum_{i=0}^{L-1} \sum_{j=i+1}^{L-1} 2 \Re{c_i^* c_j \bra{0} V^\dag A_i^\dag A_j V \ket{0}}.
    \end{split}
\end{equation}

Here, the third equality in (\ref{eq:glob_den1}) is obtained by utilizing the property
\begin{equation}
    z + \overline{z} = 2\Re{z} \quad \forall z\in\C,
\end{equation}
whereas the second equality is possible due to the following:
\begin{equation}
    \begin{split}
        \sum_{i=0}^{L-1} &\sum_{j=0}^{i-1} c_i^* c_j \bra{0} V^\dag A_i^\dag A_j V \ket{0} \\
        &= \sum_{j=0}^{L-1} \sum_{i=j+1}^{L-1} c_i^* c_j \bra{0} V^\dag A_i^\dag A_j V \ket{0} \\
        &= \sum_{i=0}^{L-1} \sum_{j=i+1}^{L-1} c_j^* c_i \bra{0} V^\dag A_j^\dag A_i V \ket{0} \\
        &= \sum_{i=0}^{L-1} \sum_{j=i+1}^{L-1} \overline{c_i^* c_j \bra{0} V^\dag A_i^\dag A_j V \ket{0}},
    \end{split}
\end{equation}
where we have used a change of index and the relation
\begin{equation}
    \label{eq:conjugate_expectation}
    \bra{\phi} U^\dag \ket{\psi} = \overline{\bra{\psi} U \ket{\phi}},
\end{equation}
with the special case where $\phi = \psi$.

Furthermore, if $c_i \in \mathbb{R}$ for all $i=0, ..., L-1$, we can further simplify and obtain (\ref{eq:real_glob_den}).

The term $\abs{\braket{b}{\psi}}^2$ of the global cost function can be computed as follows:
\begin{equation}
    \begin{split}
        \abs{\braket{b}{\psi}}^2 
        &= \abs{\bra{0} U^\dag A V \ket{0}}^2
        = \bra{0} U^\dag A V \ket{0} \overline{\bra{0} U^\dag A V \ket{0}} \\        
        &= \sum_{i=0}^{L-1} \sum_{j=0}^{L-1} c_i c_j^* \bra{0} U^\dag A_i V \ket{0} \overline{\bra{0} U^\dag A_j V \ket{0}}. \\
    \end{split}
\end{equation}

We can again split the sum and obtain
\begin{equation}
    \label{eq:glob_num1}
    \begin{split}
        \abs{\braket{b}{\psi}}^2 
        &= \sum_{i=0}^{L-1} c_i c_i^* \bra{0} U^\dag A_i V \ket{0} \overline{\bra{0} U^\dag A_i V \ket{0}} \\
        &\quad+ \sum_{i=0}^{L-1} \sum_{j=i+1}^{L-1} c_i c_j^* \bra{0} U^\dag A_i V \ket{0} \overline{\bra{0} U^\dag A_j V \ket{0}} \\
        &\quad+ \sum_{i=0}^{L-1} \sum_{j=0}^{i-1} c_i c_j^* \bra{0} U^\dag A_i V \ket{0} \overline{\bra{0} U^\dag A_j V \ket{0}}\\
        &= \sum_{i=0}^{L-1} \abs{c_i}^2 \abs{\bra{0} U^\dag A_i V \ket{0}}^2 \\
        &\quad+ \sum_{i=0}^{L-1} \sum_{j=i+1}^{L-1} c_i c_j^* \bra{0} U^\dag A_i V \ket{0} \overline{\bra{0} U^\dag A_j V \ket{0}} \\
        &\quad+ \sum_{i=0}^{L-1} \sum_{j=i+1}^{L-1}\overline{c_i c_j^* \bra{0} U^\dag A_i V \ket{0} \overline{\bra{0} U^\dag A_j V \ket{0}}} \\
        &= \sum_{i=0}^{L-1} \abs{c_i}^2 \abs{\bra{0} U^\dag A_i V \ket{0}}^2 \\
        &\quad+ \sum_{i=0}^{L-1} \sum_{j=i+1}^{L-1} 2 \Re{c_i c_j^* \bra{0} U^\dag A_i V \ket{0} \overline{\bra{0} U^\dag A_j V \ket{0}}}. \\
    \end{split}
\end{equation}

The second equality in (\ref{eq:glob_num1}) is obtained considering a change of index and (\ref{eq:conjugate_expectation}):
\begin{equation}
    \begin{split}
        \sum_{i=0}^{L-1} &\sum_{j=0}^{i-1} c_i c_j^* \bra{0} U^\dag A_i V \ket{0} \overline{\bra{0} U^\dag A_j V \ket{0}} \\
        &= \sum_{j=0}^{L-1} \sum_{i=j+1}^{L-1} c_i c_j^* \bra{0} U^\dag A_i V \ket{0} \overline{\bra{0} U^\dag A_j V \ket{0}} \\
        &= \sum_{i=0}^{L-1} \sum_{j=i+1}^{L-1} c_j c_i^* \bra{0} U^\dag A_j V \ket{0} \overline{\bra{0} U^\dag A_i V \ket{0}} \\
        &= \sum_{i=0}^{L-1} \sum_{j=i+1}^{L-1} \overline{c_i c_j^* \bra{0} U^\dag A_i V \ket{0} \overline{\bra{0} U^\dag A_j V \ket{0}}}. 
    \end{split}
\end{equation}

Finally, if $c_i\in\R$ $\forall i=0,...,L-1$, we can further simplify and obtain (\ref{eq:real_glob_num}).

\begin{figure}
\centering
\caption{Hadamard test circuit for computing the denominator in global and local cost function.\label{fig:hadamard_test_den}}
    \begin{quantikz}[wire types={q,q}]
        \lstick{$\ket{0}$}      & \gate{H} & & \ctrl{1} & \ctrl{1} & \gate{H} & \meter{} \\
        \lstick{$\ket{0}_n$} & \qwbundle{n} & \gate{V} & \gate{A_j^{\dag}} & \gate{A_i^\dag} & &
    \end{quantikz}
\end{figure}
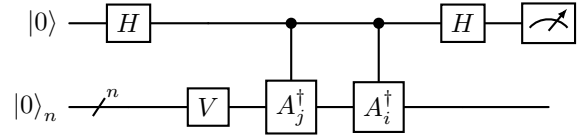

\begin{figure}
\centering
\caption{Hadamard test circuit for computing the numerator in the global cost function.\label{fig:hadamard_test_glob_num}}
    \begin{quantikz}[wire types={q,q}]
        \lstick{$\ket{0}$}      & \gate{H} & \ctrl{1} & \ctrl{1} & \ctrl{1} & \gate{H} & \meter{} \\
        \lstick{$\ket{0}_n$}  & \qwbundle{n}  & \gate{V} & \gate{A_i} & \gate{U^\dag} & &
    \end{quantikz}
\end{figure}
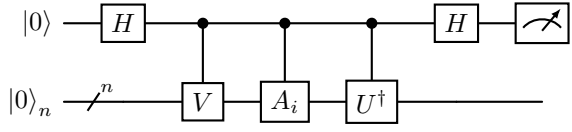

\paragraph{Local Cost Function}

The procedure for deriving the expressions to compute the local cost function (\ref{eq:local_cost_bis}) is analogous to the procedure for obtaining the expressions for the global cost function reported in the previous paragraph.
In particular, given the LCU decomposition (\ref{eq:LCU_decomposition}) of the matrix $A$, $\braket{\psi}{\psi}$ still can be computed using (\ref{eq:glob_den1}) and (\ref{eq:real_glob_den}), whereas each term $\bra{0} V^\dag A^\dag U Z_q U^\dag A V \ket{0}$ can be expressed as follows:
\begin{equation}
    \begin{split}
        \bra{0} &V^\dag A^\dag U Z_q U^\dag A V \ket{0}
        =\sum_{i=0}^{L-1} \sum_{j=0}^{L-1} c_i^* c_j \bra{0} V^\dag A_i^\dag U Z_q U^\dag A_j V \ket{0}.
    \end{split}
\end{equation}

This expression can be further simplified using a computational procedure analogous to that employed for the global cost function. By splitting the sum, introducing a change of index, and utilizing (\ref{eq:conjugate_expectation}), the result is
\begin{equation}
    \begin{split}
        \bra{0} &V^\dag A^\dag U Z_q U^\dag A V \ket{0}
        = \sum_{i=0}^{L-1} \abs{c_i}^2 \bra{0} V^\dag A_i^\dag U Z_q U^\dag A_i V \ket{0} \\
        &+ \sum_{i=0}^{L-1} \sum_{j=i+1}^{L-1} 2\Re{c_i^* c_j \bra{0} V^\dag A_i^\dag U Z_q U^\dag A_j V \ket{0}}.
    \end{split}
\end{equation}

In the case where the coefficients $c_i$ are real, the expression simplifies to (\ref{eq:real_loc_num}).

\paragraph{Real Nature of Coefficients in Pauli Decomposition of Hermitian Matrices}

Particularly noteworthy is the formulation of a necessary and sufficient condition for determining the real nature of the coefficients in the Pauli decomposition. This condition is given by the following:
\begin{proposition}
\label{real_ci_prop}
Given the LCU decomposition (\ref{eq:LCU_decomposition}) of the matrix $A$ in a Pauli basis (\ref{eq:pauli_LCU}), we have:
    \begin{equation}
        A \text{ Hermitian} \iff c_i\in\R \; \forall i=0, ..., L-1.
    \end{equation}
\end{proposition}
Since the matrices of the problem instances used in our experiments are Hermitian, Proposition \ref{real_ci_prop} guarantees that the coefficients $c_i$ are real-valued. Consequently, it is possible to use equations (\ref{eq:real_glob_den}), (\ref{eq:real_glob_num}), and (\ref{eq:real_loc_num}) to estimate the cost function values.

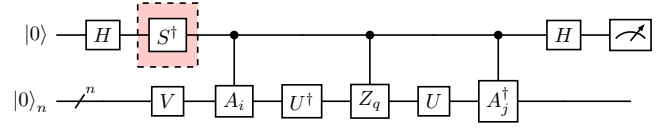
\begin{figure}
\centering
\caption{Hadamard test circuit for computing the real and imaginary parts of the expectations in the numerator of the local cost function. The $S^\dag$ gate (in the red dashed box) should be included only when estimating the imaginary part of the expectation.\label{fig:hadamard_test_loc_num}}
\resizebox{\columnwidth}{!}{
    \begin{quantikz}[wire types={q,q}]
        \lstick{$\ket{0}$}      & \gate{H} & \gate{S^\dag}\gategroup[1,steps=1,style={dashed, fill=red!20, inner xsep=2pt},background]{} & \ctrl{1}  & & \ctrl{1} & & \ctrl{1} & \gate{H} & \meter{} \\
        \lstick{$\ket{0}_n$}  & \qwbundle{n}  & \gate{V} & \gate{A_i} & \gate{U^\dag} & \gate{Z_q} & \gate{U} & \gate{A_j^\dag} & &
    \end{quantikz}
}
\end{figure}

\subsection{Circuit Executions} \label{circuit_execution_appendix}

In this section, we outline the procedure for estimating the global and the local cost function and provide insights into the associated circuit executions required.

To compute the value of the global cost function, we need to estimate the terms in the summations in (\ref{eq:real_glob_den}) and (\ref{eq:real_glob_num}). Similarly, for the estimation of the local cost function, one must find the terms in (\ref{eq:real_glob_den}) and (\ref{eq:real_loc_num}). These estimations can be obtained through Hadamard tests, as explained in Appendix \ref{hadamard_test_appendix}. Specifically, the terms in (\ref{eq:real_glob_den}) and (\ref{eq:real_loc_num}) represent real parts of expectations of unitary operators. Therefore, executing the Hadamard test with the circuits illustrated in Fig. \ref{fig:hadamard_test_den} and \ref{fig:hadamard_test_glob_num} allows to estimate these quantities.
Equation (\ref{eq:real_glob_num}) requires the estimation of both the real and imaginary parts of the expectations, which can be accomplished through the circuits in Fig. \ref{fig:hadamard_test_loc_num}. Specifically, the circuit without the $S^\dag$ gate within the red dashed box is employed for estimating the real part, while the circuit including such gate is used for the imaginary part.

Finally, we provide some insights into the number of circuit executions required for the cost function estimation.
It is crucial to recognize that, at each optimization step (with fixed parameters), the estimation of the cost function demands a specific number of Hadamard tests, depending on the number of terms in the LCU decomposition (\ref{eq:LCU_decomposition}).

Denoting by $L$ the number of matrices in the LCU decomposition, $\frac{L(L-1)}{2}$ Hadamard tests are needed for (\ref{eq:real_glob_den}), and $2L$ Hadamard tests are required for (\ref{eq:real_glob_num}). In the case of the local cost function, the estimation of (\ref{eq:real_glob_den}) still involves $\frac{L(L-1)}{2}$ Hadamard tests, whereas (\ref{eq:real_loc_num}) necessitates $n \cdot \frac{L(L-1)}{2}$ Hadamard tests. Consequently, the asymptotic count of Hadamard tests for the local cost function is approximately $n$ times the count required by the global cost function.
Therefore, to prevent excessively long computational times, it is mandatory to have a low number $L$ of matrices in the LCU decomposition.

\end{document}